\documentclass[aps,superscriptaddress,nofootinbib,eqsecnum,prd,notitlepage,twocolumn]{revtex4} 

\pdfoutput=1

\usepackage{amssymb}   
\usepackage{amsmath}   
\usepackage{graphicx}  
\usepackage{graphicx}
\usepackage{subfigure}
\usepackage{hyperref}
\usepackage{float}
\usepackage{color}

\begin{document}

%%%%%%%%LAST CHANGE BY: %%%%%%%%%%%%%%
%%% MICOL	: 26/10/2016  
%%% RUDNEI	: 27/10/2016
%%%%%%%%%%%%%%%%%%%%%%%%%%%%%%%%%%%%%%

\title{Warm inflation dissipative effects: predictions and 
constraints from the Planck data}

\author{Micol Benetti}
\email{micolbenetti@on.br}
\affiliation{Observat\'orio Nacional, 20921-400 Rio
de Janeiro, RJ, Brazil}

\author{Rudnei O. Ramos}
\email{rudnei@uerj.br}
\affiliation{Departamento de F\'{\i}sica Te\'orica, 
Universidade do Estado do Rio de Janeiro, 20550-013 Rio
de Janeiro, RJ, Brazil}

\begin{abstract}

We explore the warm inflation scenario theoretical predictions looking
at two different dissipative regimes for several representative primordial potentials. As it is well known, the warm
inflation is able to decrease the tensor-to-scalar ratio value,
rehabilitating several primordial potential ruled out in the cold
inflation context by the recent cosmic microwave background data. Here
we show that it is also able to produce a running of the running
$n_s''$ positive and within the Planck data limits. This is very
remarkable since the standard cold inflation model is unable to
justify the current indication of a positive constraint on $n_s''$.
We achieve a parameterization for the primordial power spectrum able
to take into account higher order effects as the running of the spectral index and the running of the running, and we perform  statistical
analysis using the most up-to-date Planck data to constrain the
dissipative effects. We find that the warm inflation can explain the
current observables with a good statistical significance, even for
those potentials ruled out in the simplest cold inflation scenario.

\end{abstract}


\maketitle

%%%%%%%%%%%%%%%%%%%%%%%%%%%%%%%%%%%%%%%%%%%%%%%%%%%%
\section{Introduction}

The most recent measurements of the cosmic microwave background (CMB)
from the Planck
satellite~\cite{Aghanim:2015xee,Planck2015,Ade:2015ava} are in
excellent agreement  with the assumption of adiabatic primordial
scalar perturbation with a nearly gaussian and quasi-invariant
primordial power spectrum. %and are also consistent with standard cosmological inflation~
At the same time, the observations support the standard cosmological
inflation~\cite{Guth:1980zm,Sato:1981ds,Albrecht:1982wi,Linde:1981mu,Linde:1983gd}
and the accuracy of the current data provides narrow constraints on
cosmological parameters, ruling out  large classes of models
(e.g. those predicting a large tensor-to-scalar ratio, or
large non-Gaussianities are currently discarded). 
Several inflationary primordial potentials are in agreement with data and many
other can be reconciled with the observations when the dynamics
involved during inflation go beyond %what is assumed in 
the simplest traditional cold inflation (CI) scenario.  This is the case of the warm
inflation (WI) picture~\cite{Berera:1995ie} where the presence of
nontrivial dynamics, accounting for both dissipative and related
stochastic effects, cause %yield 
valuable changes on the usual observational
quantities like in the tensor-to-scalar ratio, $r$,  the spectral
index, $n_s$, and the non-Gaussianity parameter, $f_{\rm NL}$ 
(for a representative list of recent references 
see Refs.~\cite{Bartrum:2013fia,Bartrum:2013oka,Bastero-Gil:2014raa,Ramos:2013nsa,Bastero-Gil:2014oga,Herrera:2014mca,Panotopoulos:2015qwa,Bastero-Gil:2015nja,Herrera:2015aja,Visinelli:2016rhn,Zhang:2015zta}).
In this way, some classes of inflaton potentials excluded in the CI
context by the data can be rehabilitated in the WI context,
as the monomial chaotic potential $\lambda\phi^4$~\cite{Bartrum:2013fia}.
 
The CI and the WI pictures show several peculiar differences that lead
to the above mentioned predictions.
{}For instance, in CI the
dynamics is assumed independent from the coupling of the inflaton
field $\phi$ with other fields.  However, the inflaton interaction
with other field degrees of freedom become relevant at the end of
inflation, at the time of (pre)reheating, where the energy density
stored in the inflaton is released in radiation form through decay
processes (reheating) and/or complex nonlinear dynamics (preheating).
In contrast with this picture, in the WI scenario the coupling between
the inflaton and  other fields might be strong enough to lead to  a
non-negligible radiation production rate, yet preserving the required
flatness of the inflaton potential.  The radiation production during WI
can be sufficient enough to compensate the typical supercooling
observed in CI, thus bringing to a non-isentropic inflationary
expansion, and can effectively produce a quasi-equilibrium thermal
radiation bath,  leading to a smooth transition from the inflationary
accelerated expansion to the radiation phase (for reviews, see
Refs.~\cite{Berera:2008ar,BasteroGil:2009ec}).  Dissipative and
stochastic processes typically involved in the WI
dynamics~\cite{ng1,ng2,Liu:2014ifa,Ramos:2013nsa,Bartrum:2013fia,Bastero-Gil:2014jsa,Bastero-Gil:2014raa,Vicente:2015hga}
are able to produce a strongly modified dynamics, both at the
background and the  fluctuations levels, leading to many distinguished
predictions with respect to the CI scenario.  In particular, in WI the 
primary source of density fluctuations comes
from thermal fluctuations originated in the radiation bath and
transported to the  inflaton field as adiabatic curvature
perturbations~\cite{Taylor:2000ze,Hall:2003zp},  while in CI the
density perturbations are due to quantum fluctuations of the inflaton
field~\cite{lyth2009primordial}.

In WI, the background and the inflaton fluctuations dynamics are
modified due to an extra friction term $\Upsilon
\dot{\phi}$, that accounts for the energy transfer between the
inflaton and the radiation. The specific form for the dissipation
coefficient $\Upsilon$ depends on  the model building details, as the
field interactions and the parameters regime.  A particular form of
dissipation coefficient, highly studied in the literature~\cite{Berera:2008ar,Bastero-Gil2011,BasteroGil:2012cm}, shows a cubic dependence with the thermal radiation bath temperature, $\Upsilon \propto T^3$. 
Other forms, obtained depending on the scheme
and the interactions form of the inflaton field with other field
degrees of freedom, are also possible. {}For instance, the form
$\Upsilon \propto 1/T$, was mainly considered in the first works of WI.
However, the dynamical regime where this type of dissipation coefficient emerges proved to be troublesome, 
since it happens in a regime where large thermal corrections can be produced, 
affecting the inflaton potential and  precluding WI from happening in the simplest models~\cite{Berera:1998gx,Yokoyama:1998ju}
(see, however, Ref.~\cite{Berera:1998px} for a realization of WI
for a specific model in this regime).  
Recently, in Ref.~\cite{Bastero-Gil:2016qru} it was shown that
an inflation model making use of the  collective symmetry breaking of the
Little Higgs type of models~\cite{Schmaltz:2005ky}, that leads to a dissipation coefficient $\Upsilon \propto T$, 
can resolve the objections posed in the earlier
references~\cite{Berera:1998gx,Yokoyama:1998ju} and resulting in a successful WI realization.

In this work we provide an extended analysis on the predictions  of
the two successful dissipation forms in WI %obtained up to this date, i.e.,  
$\Upsilon \propto T^3$ and $\Upsilon \propto T$.  We focus on
key parameters like  $r$ and $n_s$, as well as the running of the
spectral index $n_s'$ and the running of the running $n_s''$, and we
also give results for the tensor spectral index $n_t$.  To our
knowledge, this is the first work that studies  predictions and
constraints in WI of the running of the running.  This parameter has
recently attracted increasing
attention~\cite{Escudero:2015wba,Cabass:2016ldu,vandeBruck:2016rfv}
due to the Planck results hinting on a rather larger and positive
value for $n_s''$ at $2\sigma$-confidence level (CL)~\cite{Planck2015}.
Interestingly, while standard CI models can only predict a very small
and {\it negative} $n_s''$,  here we show that WI can lead to a {\it
  positive} $n_s''$, depending of the inflaton potential and
dissipative regime.  We achieve our predictions studying a variety of
representative potentials for the inflaton, covering both large and
small field models, extending earlier works that made use primarily of
the chaotic quartic and  hilltop type of inflaton potentials.  We
finally perform a statistical analysis for these
inflationary models in  the considered dissipation regimes, providing
the most up-to-date analysis for the WI.  
Our results provide the essential link between the (many) aspects studied 
in the literature and the precise measurements of the CMB radiation anisotropies.
{}Furthermore, our results place a greater demand not only on better constrain the existing WI models, 
but also on future model building.

This work is organized as follows.  In
Sec.~\ref{sec:Models} we introduce the basic equations describing the
dynamics of WI and give the scalar curvature power spectrum. The formulas 
for the primordial observables as the tensor-to-scalar ratio, the spectral 
index, the running of the spectral index and the running of the 
running are presented. We also introduce the
two dissipation regimes that we are going to analyze in this work.  In
Sec.~\ref{sec:observables} we show the predictions of the primordial
observables for several inflationary potentials.  At the same time, it
is determined the behavior for these quantities as a function of the
strength of the dissipation in a warm scenario.  In
Sec.~\ref{sec:analysis and results} we introduce our analysis method
as well the tools and the data sets used and we discuss the obtained
results.  {}Finally,  in Sec.~\ref{sec:Conclusions} we present our
conclusions.

%%%%%%%%%%%%%%%%%%%%%%%%%%%%%%%%%%%%%%%%%%%%%%%%%%%%
\section{The warm inflation picture} 
\label{sec:Models}

The WI dynamics is characterized by the coupled system of the
background equation of motion for the inflaton field, $\phi(t)$,  the
evolution equation for the radiation energy density, $\rho_R(t)$, and
the {}Friedmann equation given, respectively, by 

\begin{equation} \label{eom_inflation_back}
\ddot{\phi}(t) + 3H(1+Q)\dot{\phi}(t) + V_{,\phi} = 0,
\end{equation}

\begin{equation} \label{eom_radiation_back}
\dot{\rho}_R + 4H\rho_R  = 3H Q \dot{\phi}^2,
\end{equation}

\begin{equation}
H^2 = \frac{1}{3 M_P^2} \left( \rho_\phi + \rho_R \right),
\end{equation}
where the dissipation ratio $Q$ is defined as $Q \equiv \Upsilon/(3
H)$ with $\Upsilon$ the dissipation coefficient in WI,  $H$ is the
Hubble factor,  $V$($\phi$) is the primordial inflaton potential,  $\rho_\phi
= {\dot\phi}^2/2 + V(\phi)$ is the inflaton energy density and
$M_P\equiv 1/\sqrt{8 \pi G}=2.4 \times 10^{18}$ GeV is the reduced
Planck mass.  {}For a radiation bath of relativistic particles, the radiation energy density is
%we have that 
$\rho_R=\pi^2 g_* T^4/30$, where $g_*$ is the effective
number of light degrees of freedom  ($g_*$ is fixed according to 
the dissipation regime and interactions form used in WI).

In the slow-roll regime, Eqs.~(\ref{eom_inflation_back}) and
(\ref{eom_radiation_back}) can be approximated to

\begin{align} \label{eoms_sra}
 3H(1+Q)\dot{\phi} &\simeq - V_{,\phi} ,  \\
 \label{eoms_sra2}
 \rho_R  &\simeq  \frac{3}{4}Q\dot{\phi}^2 ,
\end{align}
and the slow-roll conditions in the WI given
by~\cite{Berera:2008ar,BasteroGil:2009ec}

\begin{align} 
\varepsilon_\phi &= \frac{M_p^2}{2} \left( \frac{V_{,\phi}}{V}
\right)^2 < 1+Q ,
\label{varepsilon} 
\\  \eta_\phi &= M_p^2  \frac{V_{,\phi\phi}}{V}  < 1+Q ,
\label{eta}
 \\  \beta_\phi & = M_p^2 \frac{\Upsilon_{,\phi} V_{,\phi}}{\Upsilon
   V}   < 1+Q .
\end{align}

%%%%%%%%%%%%%%%%%%%%%%%%%%%%%%%%%%%%%%%%%%%%%%%%%%%
\subsection{The dissipation coefficient}

The dissipation coefficient $\Upsilon$ embodies the microscopic
physics resulting from the interactions between the inflaton and  the
other fields that can be present,  taking into account the different
dissipative processes arising from these
occurrences~\cite{Berera:2008ar,Bastero-Gil2011}.  {}For instance,
most of the WI models make use of a structure of interactions such
that the inflaton is coupled to heavy intermediate fields, that are in
turn coupled to light radiation fields. 
As the inflaton slowly moves according to its potential, it can
trigger the decay of these heavy intermediate fields 
into the light radiation fields 
and generates a dissipation term for the
inflaton~\cite{Berera:2002sp}. In this case, the resulting dissipation
coefficient can be well described by the
expression~\cite{Berera:2008ar,Bastero-Gil2011,BasteroGil:2012cm}

\begin{equation}\label{upsilon}
\Upsilon_{\rm cubic} = C_{\phi} \frac{T^3}{\phi^2}, 
\end{equation}
where $C_{\phi}$ is a dimensionless parameter that depends on the
interactions specifics.   Hereafter we refer to the above
$\Upsilon_{\rm cubic}$ as the {\it cubic dissipation coefficient}.
This is obtained in the so-called {\it low temperature } regime for
WI~\cite{Berera:2008ar,Bastero-Gil2011,BasteroGil:2012cm}, where the
inflaton only couples to the heavy intermediate fields, whose masses
are larger than the radiation temperature and, thus, the inflaton gets
decoupled from the radiation fields.

{}For the cubic dissipation coefficient we can find explicit
expressions for the evolutions of $Q$, $T/H$ and $\phi$ as a function
of the number of e-folds $N_e$ and in the slow-roll approximation,
given by~\cite{BasteroGil:2009ec}

\begin{eqnarray}
\!\!\!\!\!\frac{d\phi}{dN_e}&=&-\frac{\phi \, \sigma}{1+Q}, 
\label{dphidN}
\\   \!\!\!\!\!\frac{dQ}{dN_e}&=&\!\frac{Q}{1+7Q}(10\varepsilon_\phi -
6\eta_\phi + 8\sigma), 
\label{dQdN}
\\   \!\!\!\!\frac{d(\frac{T}{H})}{dN_e} \!\!&=&\!\! \frac{2\,
  \frac{T}{H}}{1+7Q}\left(\! \frac{2+4Q}{1+Q}\varepsilon_\phi -
\eta_\phi + \frac{1-Q}{1+Q}\sigma \!\right), 
\label{dnudN}
\end{eqnarray}
where we have defined $\sigma =M_P^2 V_{,\phi}/(\phi V)$.

More recently, it was realized another mechanism able to lead to a
successful WI regime~\cite{Bastero-Gil:2016qru}, based on a collective
symmetry where the inflaton is a pseudo-Goldstone boson.  In this
case the inflaton can be directly coupled to the radiation fields and
gets protection from the large potential thermal corrections due to
the symmetries obeyed by the model.  The resulting dissipative
coefficient, here obtained in the {\it large temperature}
regime (where the fields coupled to the inflaton are light with respect
to the ambient temperature), is given simply by~\cite{Bastero-Gil:2016qru}

\begin{equation}\label{upsilon2}
\Upsilon_{\rm linear} = C_{\phi} T.
\end{equation}
Hereafter we refer to the above equation as the {\it linear
  dissipation coefficient}.  Also in this case, we can find analogous
expressions of Eqs.~(\ref{dphidN})-(\ref{dnudN}).  While the equation
(\ref{dphidN})  for $\phi$ remains unchanged, the equations for $Q$
and $T/H$ are now  given by~\cite{Bastero-Gil:2016qru}

\begin{align}
\frac{dQ}{dN_e} &=  \frac{Q}{3+5Q}(6\varepsilon_\phi- 2\eta_\phi), 
\label{dQdN2}
\\  \frac{d(T/H)}{dN_e} &= \frac{T/H}{3+5Q}(6\varepsilon_\phi-
2\eta_\phi).
\label{dnudN2}
\end{align}

%%%%%%%%%%%%%%%%%%%%%%%%%%%%%%%%%%%%%%%%%%%%%%%%%%%%%%%%%%%
\subsection{The primordial power spectrum in WI}

The primordial power spectrum for WI at horizon crossing has been 
studied and determined in many previous 
papers~\cite{Hall:2003zp,Graham:2009bf,BasteroGil:2011xd,Ramos:2013nsa} 
and can be written in the form
\begin{eqnarray} \label{spectrum}
\!\!\!\!\!\!\!\!\!\Delta_\mathcal{R}\! =\!\left(\!\frac{ H_{*}^2}{2
  \pi\dot{\phi}_*}\!\right)^2\!\!\left(\!1\!  +\!2n_*
\!+\!\frac{2\sqrt{3}\pi Q_*}{\sqrt{3\!+\!4\pi Q_*}}{T_*\over
  H_*}\!\right)\! G(Q_*),
\end{eqnarray}
where we indicate with a subindex ``$*$" those quantities evaluated at horizon crossing.
In the latter formula, $n_*$ denotes the inflaton statistical distribution due to the presence of the radiation bath.  Here we assume a thermal equilibrium distribution function $n_* \equiv n_{k_*}$ for the inflaton and, thus, it assumes the Bose-Einstein distribution form,
$n_* =1/[\exp(H_*/T_*) -1]$.  
The function $G(Q_*)$ in Eq.~(\ref{spectrum}) accounts for the growth of inflaton fluctuations
due to its coupling with radiation, and  can only be determined
numerically by solving the full set of perturbation equations found in
WI~\cite{Graham:2009bf,BasteroGil:2011xd,Bastero-Gil:2014jsa,Bastero-Gil:2016qru}.
According to the method of the previous works, we use a numerical
fit for $G(Q_*)$.  {}For the linear dissipation coefficient
$\Upsilon_{\rm linear}$, we get 
\begin{eqnarray} \label{growing_mode}
G_{\rm linear}(Q_*)\simeq 1+ 0.335 Q_*^{1.364}+ 0.0185Q_*^{2.315},
\end{eqnarray} 
while a similar function, appropriate for the cubic dissipation
coefficient  $\Upsilon_{\rm cubic}$, is 
\begin{eqnarray} \label{growing_mode2}
G_{\rm cubic}(Q_*)\simeq 1+ 4.981 Q_*^{1.946}+ 0.127 Q_*^{4.330}.
\end{eqnarray} 
The higher powers in the dissipation ratio of the latter formula is
due to the higher power of the temperature in Eq.~(\ref{upsilon}) with
respect to the linear form of $\Upsilon_{\rm linear}$.  This implies a
stronger coupling of the inflaton perturbations with the radiation
ones and leads to a stronger effect on the primordial power spectrum
at larger dissipation ratio
$Q$~\cite{Graham:2009bf,BasteroGil:2011xd,Bastero-Gil:2014jsa}.

As said, the quantities in the primordial power spectrum of
Eq.~(\ref{spectrum}) are evaluated when the relevant CMB modes become
super-horizon, e.g. when the modes cross the Hubble radius around $N_*
= 50 - 60$ e-folds before the end of inflation.  The precise value of
$N_*$ depends on the details of the reheating and later expansion
history after inflation. Due to the current few knowledge about this
epoch for both the CI and WI scenarios, we choose to fix $N_*$ at the
middle value of $55$ for definiteness
\footnote{This value can be justified when studying the final stages
  of the WI dynamics. At the end of WI, we typically have that the
  inflaton and radiation energy densities,  $\rho_\phi$ and $\rho_R$,
  respectivley, satisfy $\rho_\phi \approx \rho_R$ and the equation of
  state quickly evolves to that of radiation, $w \approx 1/3$,
  producing in general $N_* \approx 55$~\cite{WIreheat}.}. 

It is convenient, for later purposes, to write Eq.~(\ref{spectrum}) as  

\begin{equation} \label{Pk}
\Delta_{{\cal R}}(k/k_*) =  P_0(k/k_*) {\cal F} (k/k_*),
\end{equation}
where we have defined

\begin{equation}
P_0(k/k_*) \equiv  \left(\frac{ H_{*}^2}{2 \pi\dot{\phi}_*}\right)^2 ,
\end{equation}
and the {\it enhancement term}

\begin{equation}
{\cal F} (k/k_*) \equiv  \left(1+2n_* + \frac{2\sqrt{3}\pi
  Q_*}{\sqrt{3+4\pi Q_*}}{T_*\over H_*}\right) G(Q_*).
\label{calF}
\end{equation}
Note that the scalar spectral amplitude value at the pivot scale $k_*$
is set by the CMB data at   $\Delta_{{\cal R}}^2(k=k_*) \simeq 2.2
\times 10^{-9}$, with $k_*=0.05 {\rm Mpc}^{-1}$ as considered by the
Planck Collaboration~\cite{Planck2015}.

%%%%%%%%%%%%%%%%%%%%%%%%%%%%%%%%%%%%%%%%%%%%%%%%%%%%%%%%%%
\subsection{Observable quantities}

Given the scalar curvature power spectrum, the tensor-to-scalar ratio
$r$ and the spectral tilt $n_s$ follow from their usual definitions
just like in the CI case,

\begin{equation}
r= \frac{\Delta_{T}}{\Delta_{{\cal R}}},
\label{eq:r}
\end{equation}
and

\begin{equation}
n_s -1 = \lim_{k\to k_*}   \frac{d \ln \Delta_{{\cal R}}(k/k_*) }{d
  \ln(k/k_*) },
\label{eq:n}
\end{equation}
where $\Delta_{T} = 2 H_*^2/(\pi^2 M_p^2)$ is the tensor power
spectrum.  Due to the weakness of gravitational interactions, the
tensor modes are not affected by the dissipative dynamics and
$\Delta_{T}$ remains unaltered from the CI
result~\cite{Ramos:2013nsa}. On the other hand, the dissipative and
thermal effects modify the scalar power spectrum through the
enhancement term $ {\cal F} (k/k_*)$, producing a decrease of the
tensor-to-scalar ratio value in WI.

Looking at the latest Planck results, we quote the $\Lambda$CDM model
constraint of $n_s=0.9655\pm 0.0062$ ($68\%$ CL) using Planck TT+lowP
data~\cite{Planck2015}. {}For the extended model $\Lambda$CDM$+r$,  we
obtain $r < 0.08$ ($95\%$ CL) when the Planck TT+lowP dataset is
combined with BICEP/Keck Array data~\cite{Ade:2015tva, Planck2015} at
the pivot scale $k_* = 0.002 {\rm Mpc}^{-1}$.  

Considering a $\Lambda$CDM$+n_s'$ model, with the running of the
scalar index,

\begin{equation}
n_s'= \lim_{k \to k_*}  \frac{dn_s(k/k_*) }{d \ln(k/k_*) },
\label{dns}
\end{equation}
the data prefers a tiny negative value $n_s' = -0.0084 \pm 0.0082$
($68\%$ CL, TT+lowP), with a small improvement of the maximum
likelihood with respect to a powel-law spectrum ($\Delta\chi =
-0.8$)~\cite{Planck2015}.   The data give also the possibility of a
small running of the running,

\begin{equation}
n_s''= \lim_{k \to k_*}  \frac{d^2 n_s(k/k_*) }{d \ln(k/k_*)^2 },
\label{d2ns}
\end{equation}
and in this case, at $68\%$ CL using the TT+lowP data, the values
obtained for a $\Lambda$CDM$+n_s'+n_s''$ model are

\begin{eqnarray}
\nonumber n_s &=& 0.9569 \pm 0.0077,\\ \nonumber  n_s' &=& 0.011 \pm
0.014,\\ \nonumber  n_s'' &=& 0.029 \pm 0.015,
\end{eqnarray}
which seems to produce a better fit to the temperature spectrum at low
multipoles according to the Planck results, such that $\Delta\chi =
-4.8$~\cite{Planck2015}.   The positive constraints on $n_s''$ takes a
particularly relevance since, for single field slow-roll CI models, the
running of the running 
%(i.e.,third order in the slow-roll parameters)
is expected to be progressively smaller, and usually negative.   So,
we can see a %considerable 
tension between the current analysis  and the current favoured minimal inflation scenario (for recent discussions
on this issue, see Refs.~\cite{Escudero:2015wba,Cabass:2016ldu,vandeBruck:2016rfv}).
%
%%%%%%%%%%%%%%%%%%%%%%%%%%%%%%%%%%%%%%%%%%%%%%%%%%%%%%%%%%%%%%%
\section{Theoretical predictions in the WI context}
\label{sec:observables}
The purpose of this work is to give a clear overview of the effects of the two WI dissipation forms %$\Upsilon_{\rm cubic}$ and $\Upsilon_{\rm linear}$ 
considered in the previous section on the inflationary observables, 
% respectively, 
as well as their statistical significance
when compared with current CMB data. 
To get a more complete study, we select several representative classes of primordial potentials, namely:

\begin{enumerate}

\item {\it Chaotic Quartic Potential}:

\begin{equation}
V_{\rm quartic}(\phi) = \frac{\lambda}{4} \phi^4,
\label{quartic}
\end{equation}

\item {\it Chaotic Sextic Potential}:

\begin{equation}
V_{\rm sextic}(\phi) = \frac{\lambda}{6} \phi^6,
\label{Sextic}
\end{equation}

\item {\it Hilltop Quadratic Potential}:

\begin{equation}
V_{\rm hilltop}(\phi) = \frac{\lambda M_P^4}{2}
\left[1-\frac{\gamma}{2}\left(\frac{\phi}{M_p}\right)^{2}\right],
\label{hilltop}
\end{equation}

\item {\it Higgs-like Potential}:

\begin{equation}
V_{\rm Higgs}(\phi) = \frac{\lambda}{4} (\phi^2 - v^2)^2,
\label{higgs}
\end{equation}

\item {\it Plateau Sextic Potential}:

\begin{equation}
V_{\rm plateau}(\phi) = \frac{\lambda v^6}{12 M_P^2} \left(1-
3\frac{\phi^2}{v^2} +2 \frac{\phi^6}{v^6} \right),
\label{Plateau}
\end{equation}
\end{enumerate}
%
%%%%%%%%%%%%%%%FIGURE01%%%%%%%%%%%%%%%%%%%
\begin{center}
\begin{figure}[!Htb]
\subfigure[]{\includegraphics[width=8cm]{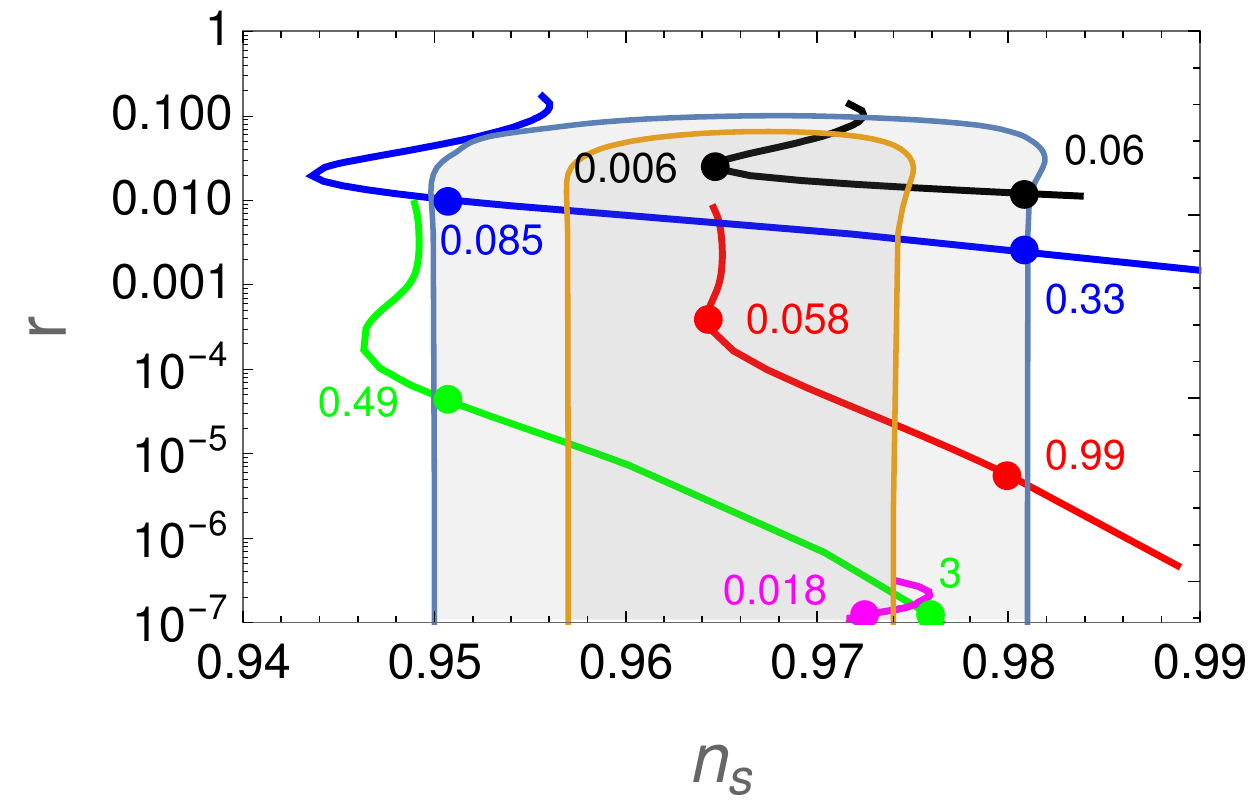}}
\subfigure[]{\includegraphics[width=8cm]{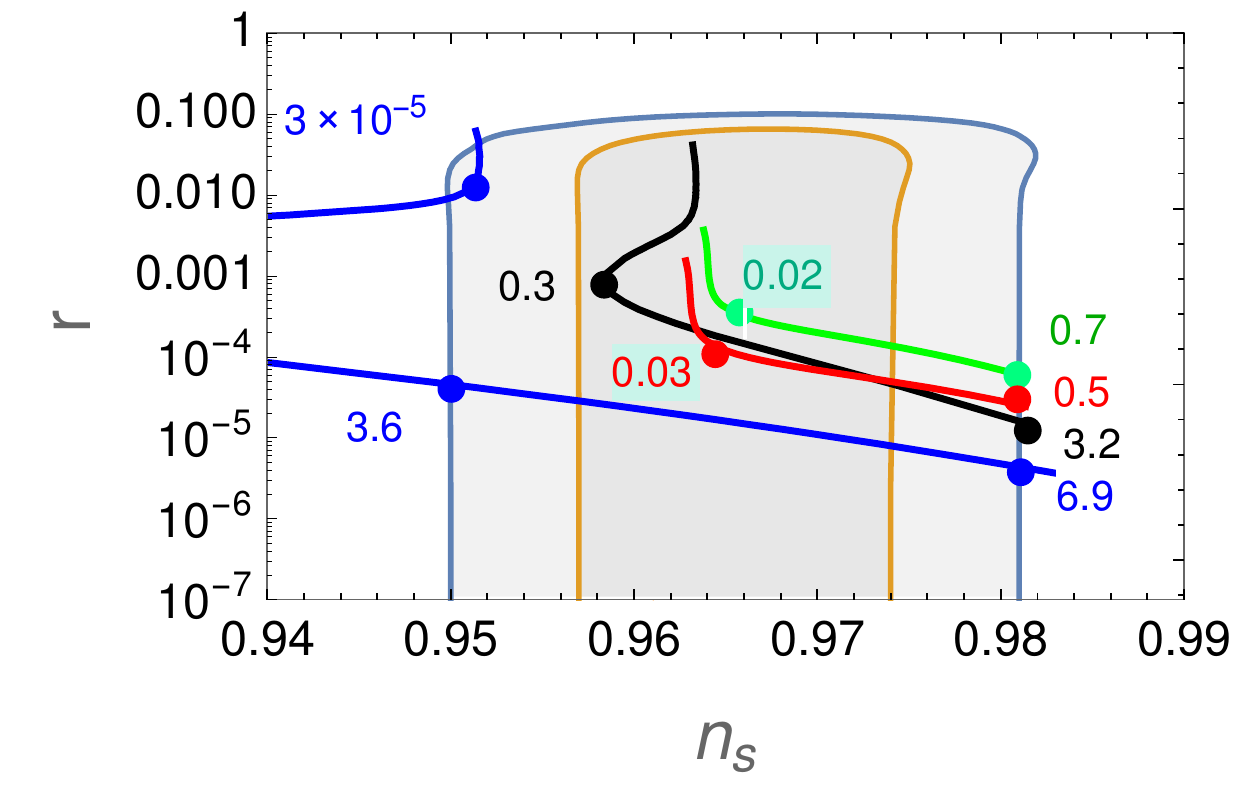}}
\caption{The spectral index $n_s$ and the tensor-to-scalar ratio $r$
  in the plane $(n_s,r)$ for different values of the dissipation ratio
  $Q_*$ (indicated by the numbers next to the curves), for the chaotic
  quartic potential (black line), hilltop (red line), Higgs potential
  (green line), chaotic sextic  potential (blue line) and for the
  plateau sextic potential (magenta line).  The contours are for the
  $68\%$ and $95\%$ C.L. results from Planck 2015. Panel (a) are the
  results for the cubic  dissipation coefficient, while panel (b)
  gives the results when the dissipation coefficient is linear. }
\label{fignsXr}
\end{figure}
\end{center}
%%%%%%%%%%%%%%%%%%%%%%%%%%%%%%%%%%%%%%%%
%
where $\lambda$ and $\gamma$ are free dimensionless constant
parameters ($\lambda$ can be fixed by the normalization condition on
the amplitude of the scalar power spectrum) and $v$ is the vacuum
expectation value (VEV) with dimension of energy.  

Both the chaotic potentials (quartic and sextic) are representative
examples of large field models, and the quartic potential is a typical
prototype potential used in many renormalizable scalar field theories.
While in the CI picture these are examples of potentials unfavorable
by the Planck data~\cite{Planck2015},  they can be rehabilitated in
the context of
WI~\cite{Bartrum:2013fia,Bastero-Gil:2016qru,Ramos:2013nsa}.
Instead, potentials like the hilltop, the Higgs and the plateau ones
are typical examples of small field models of inflation and are found to be in
agreement with the Planck data in both cold and warm inflation
pictures~\cite{Bartrum:2013oka}. 
%
%%%%%%%%%%%%%%%%%FIGURE02%%%%%%%%%%%%%%%%%%%
\begin{center}
\begin{figure}[!Htb]
\subfigure[]{\includegraphics[width=7.5cm]{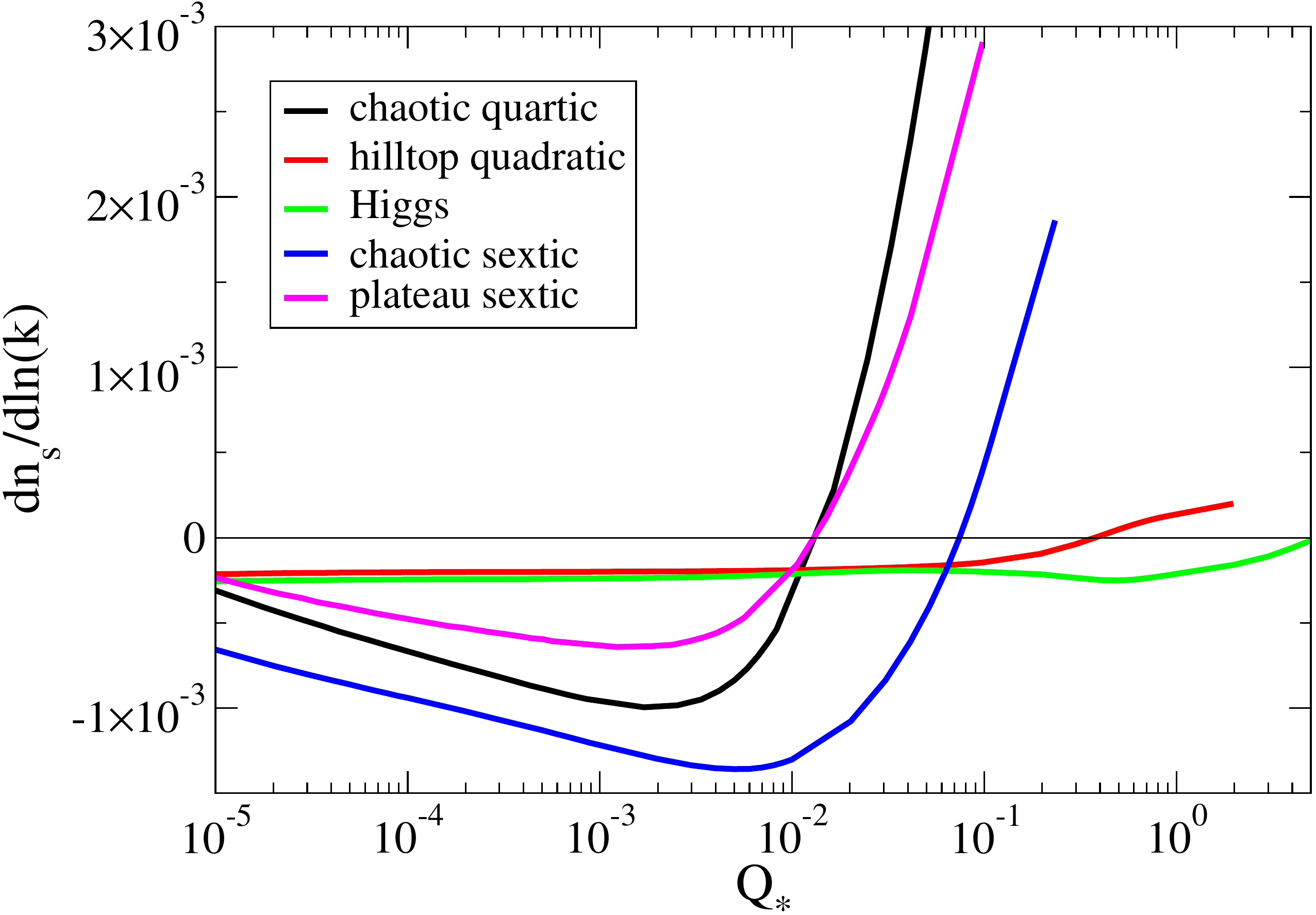}}
\subfigure[]{\includegraphics[width=7.5cm]{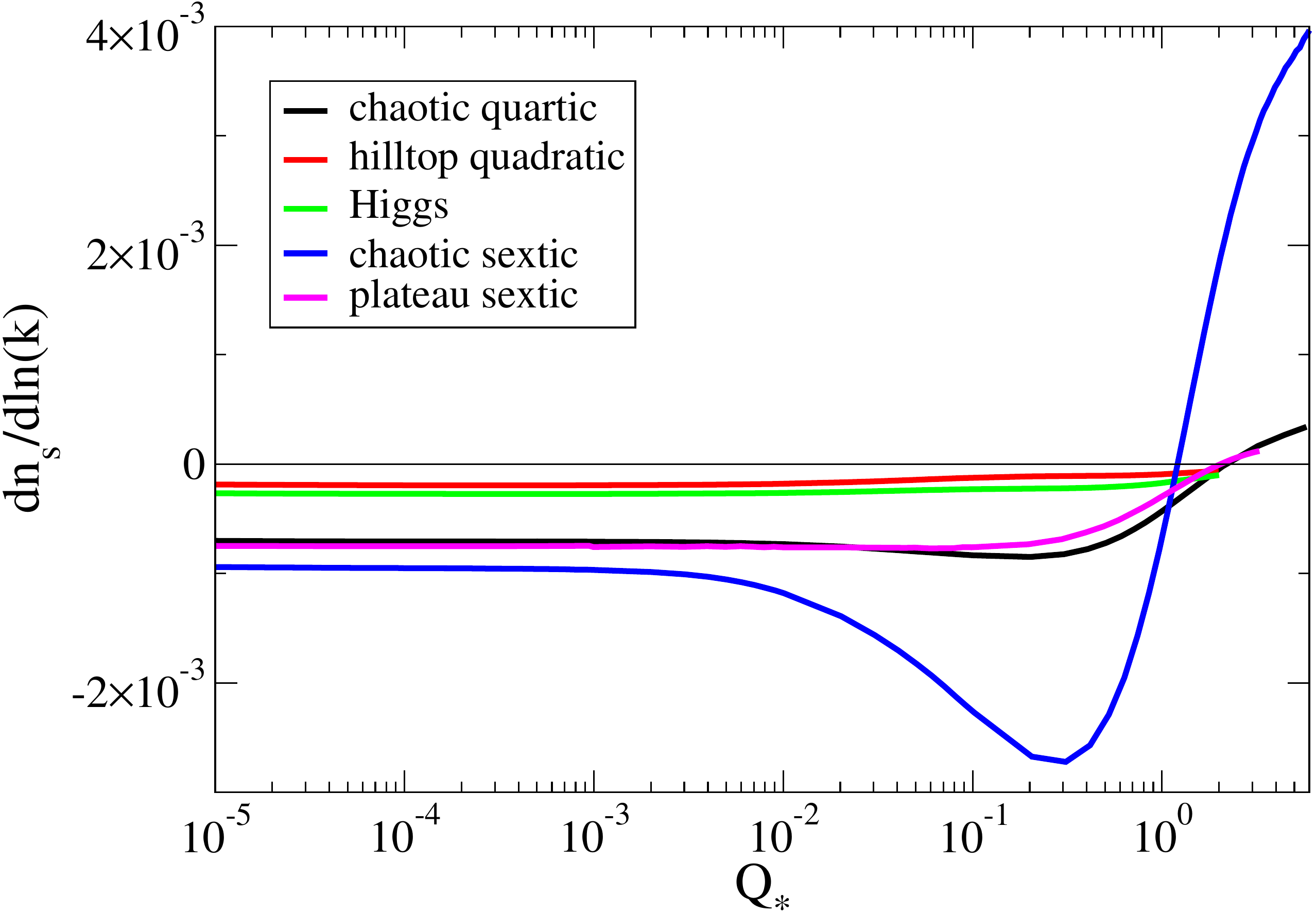}}
\caption{The running spectral index, $n_s'$, as a function of the
  dissipation ratio $Q_*$.  Panel (a) are the results for the cubic
  dissipation coefficient, while panel (b)  gives the results when the
  dissipation coefficient is linear.}
\label{figdns}
\end{figure}
\end{center}
%%%%%%%%%%%%%%%%%%%%%%%%%%%%%%%%%%%%%%%%

In order to produce $n_s$ predictions compatible with the latest CMB
data for a wider range of $Q_*$ variation, we choose different
setting values for the arbitrary parameters $\gamma$ and $v$.  Thus, the
vacuum expectation value $v$ for the plateau sextic potential
Eq.~(\ref{Plateau})  is set to $v = M_P$, while  in the Higgs
potential Eq.~(\ref{higgs})  we have assumed $v=13.2\, M_P$.  In
particular, the latter choice is a limiting value below which $n_s$ is
at the border of the values constrained from Planck in CI model, i.e.,
$n_s < 0.95$ for $v<13.2\, M_P$.  We follow the same idea to choose
the $\gamma$ value of the hilltop potential,  depending on the type of
dissipation coefficient. We assume $\gamma=0.0147$ for the cubic and
$\gamma=0.025$ for the linear form of dissipation, respectively.

The plane $(n_s,r)$ for each considered potentials and for the two
different forms of the dissipation coefficient are shown in
{}Fig.~\ref{fignsXr} for different values of $Q_*$. 
In the panel (a) %of {}Fig.~\ref{fignsXr} 
we can see the prediction when a cubic dissipation coefficient is considered,
while in the panel (b) we have the results for the linear dissipation
form.   The results for the plateau sextic potential does not appear
in the 
%%%%%%%%%%%%%%%%%FIG 3%%%%%%%%%%%%%%%%
\begin{center}
\begin{figure}[!htb]
\subfigure[]{\includegraphics[width=4cm]{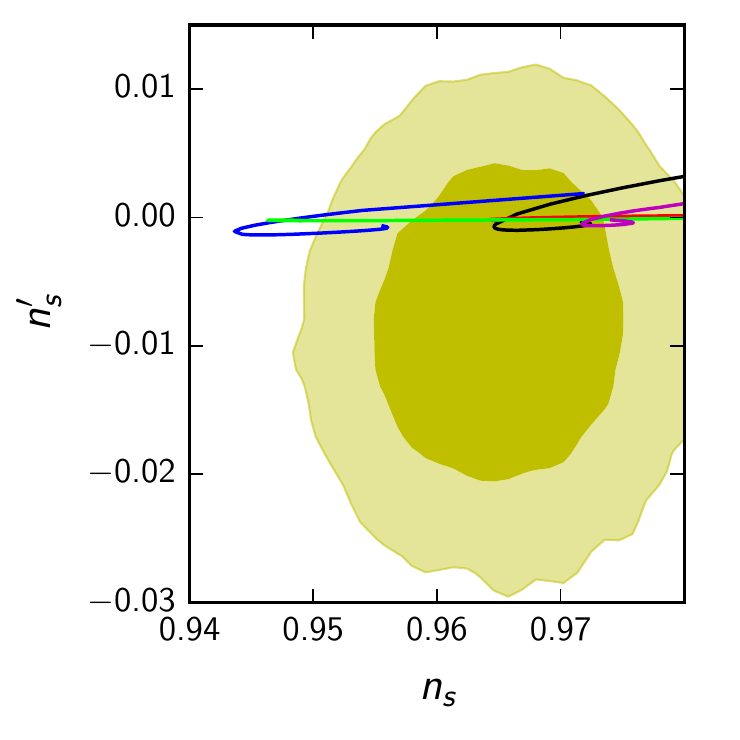}}
\subfigure[]{\includegraphics[width=4cm]{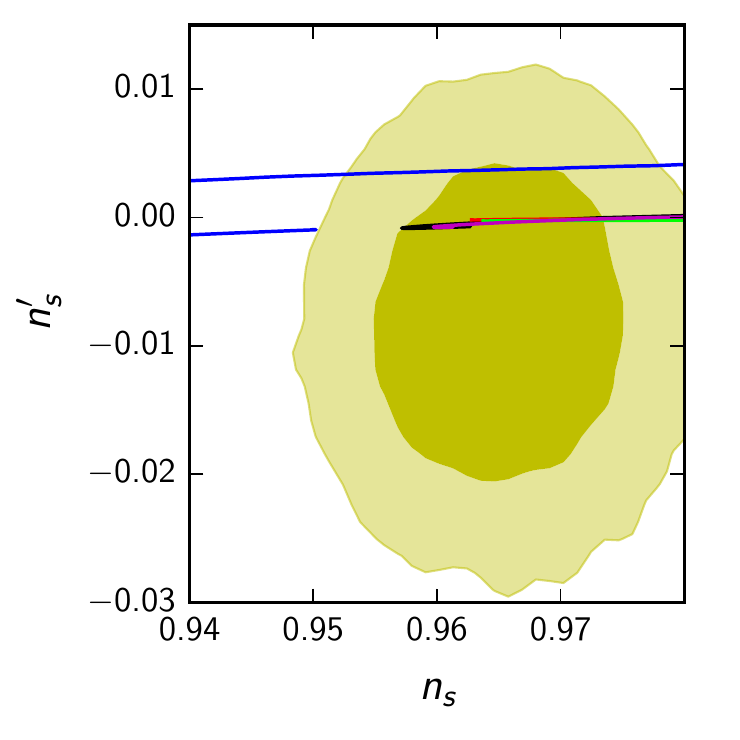}}
\subfigure[]{\includegraphics[width=4cm]{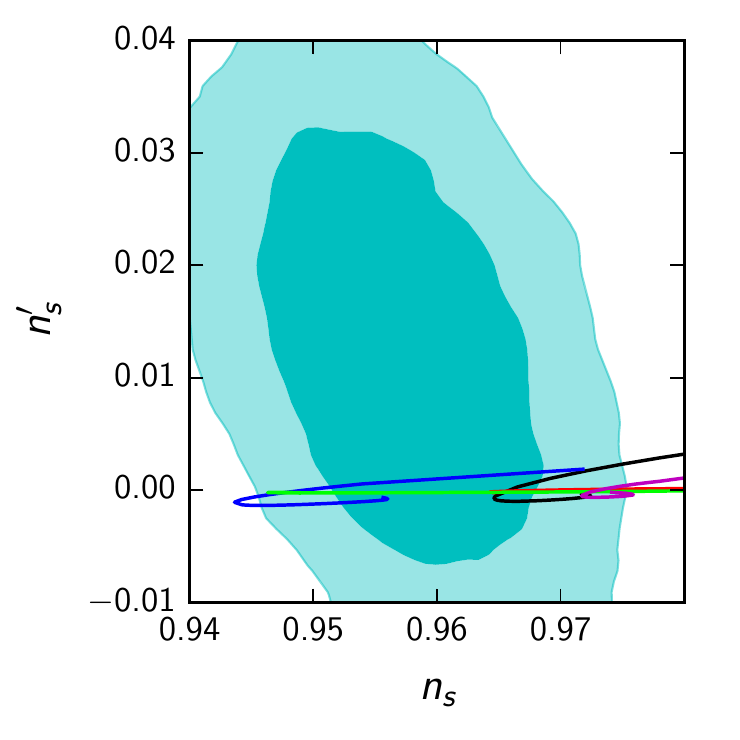}}
\subfigure[]{\includegraphics[width=4cm]{LCDM+nrun+nrunrun_2D_ns_nrun.pdf}}
\caption{Marginalized joint $68\%$ and $95\%$ C.L. for the
  $\Lambda$CDM$+n_s'$ model (yellow contours) and the
  $\Lambda$CDM$+n_s'+n_s''$ model (cyan contours) from Planck TT+lowP
  data.  The CL are compared with the WI theoretical predictions of
  the cubic dissipation regime (left column) and linear dissipation
  regime (right column).  The color schemes for the curves refer to 
  each of the inflaton potentials considered and are the same as the 
  ones used in {}Fig.~\ref{figdns}.}
\label{fig:2D_run}
\end{figure}
\end{center}
%%%%%%%%%%%%%%%%%%%%%%%%%%%%%%%%%%%%%%%%%
panel (b) since it produces values of $r$ well below the scale
shown in the figure, with $r \lesssim 10^{-8}$.  The smallest initial
value considered for the dissipation ratio $Q_*$ is always $10^{-5}$
and its value increases up to the values shown at the most right side
of the plots. %in {}Fig.~\ref{fignsXr}. 
We can see that all the curves
show a decrease in $r$ when $Q_*$ increases.  This is easily explained
recalling the Eq.~(\ref{eq:r}) and that,  while the gravitational
tensor modes remain unchanged,  the scalar power spectrum is enhanced
by the factor $ {\cal F} (k/k_*)$,  Eq.~(\ref{calF}).
%, due to the effects of dissipation and the presence of the thermal bath. 
This is why, for example, the simple chaotic quartic or sextic potentials,
which are discarded in the CI picture for their large tensor-to-scalar
prediction, can be again in accordance with the observational
data in the WI case.   We also note that the results obtained for the
linear dissipation coefficient  allow a larger range for $Q_*$ in both
the chaotic potentials, while for the hilltop and Higgs potentials we
have an opposite behavior.

In {}Fig.~\ref{figdns} we show the running of the spectral index,
$n_s'$, as a function of the dissipation ratio $Q_*$ considering the
cubic dissipation coefficient, % whose results are shown in 
panel (a),
and for the linear dissipation coefficient, %whose results are shown in 
panel (b).
We can see that the running value is always small and
within the Planck ranges for any value of $Q_*$ and for each of the
models  we have considered. Noteworthy, $n_s'$ remains negative for
%
%%%%%%%%%%%%%%%%%%FIGURE04%%%%%%%%%%%%%%%%%%%
\begin{center}
\begin{figure}[!htb]
\subfigure[]{\includegraphics[width=7.5cm]{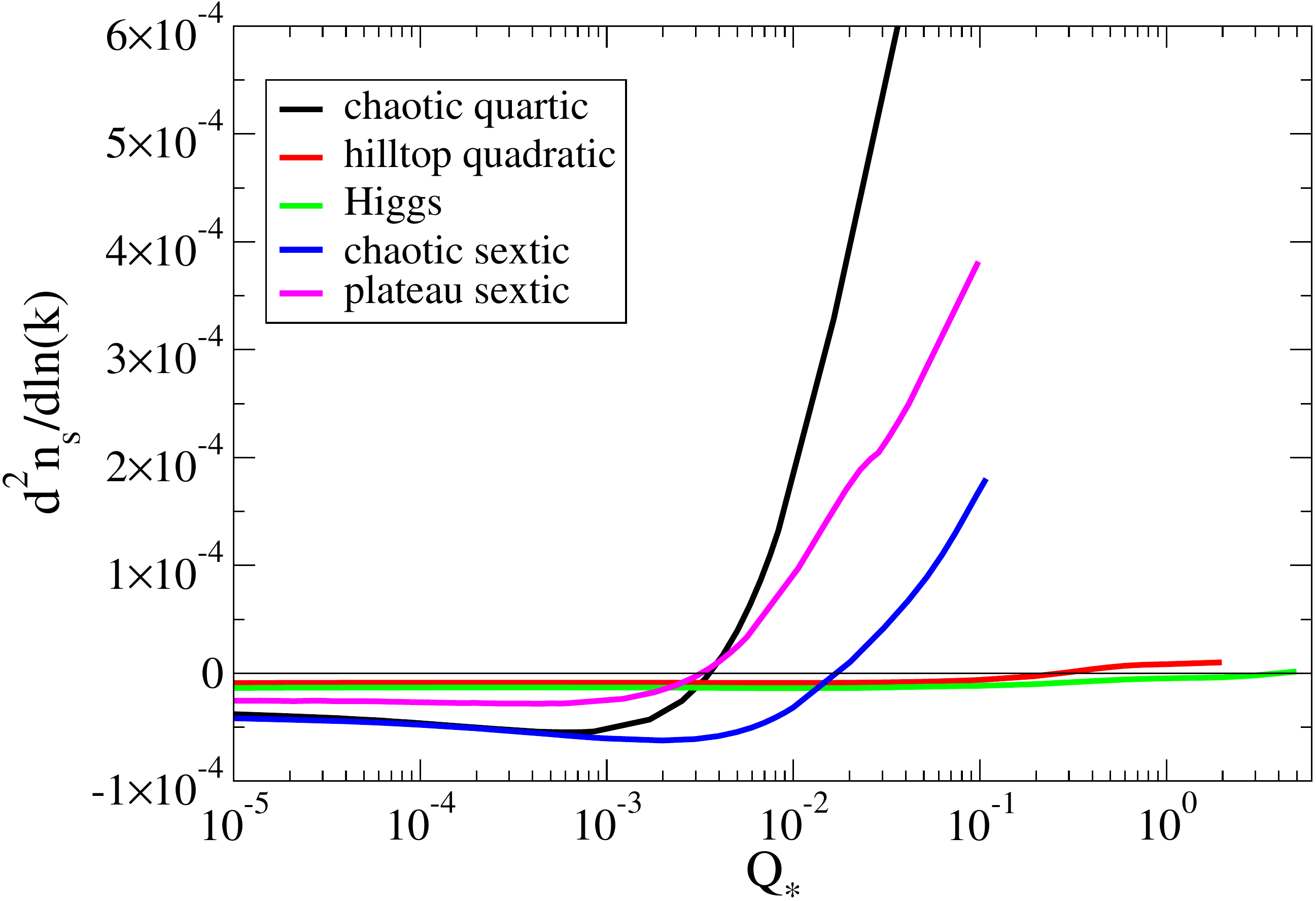}}
\subfigure[]{\includegraphics[width=7.5cm]{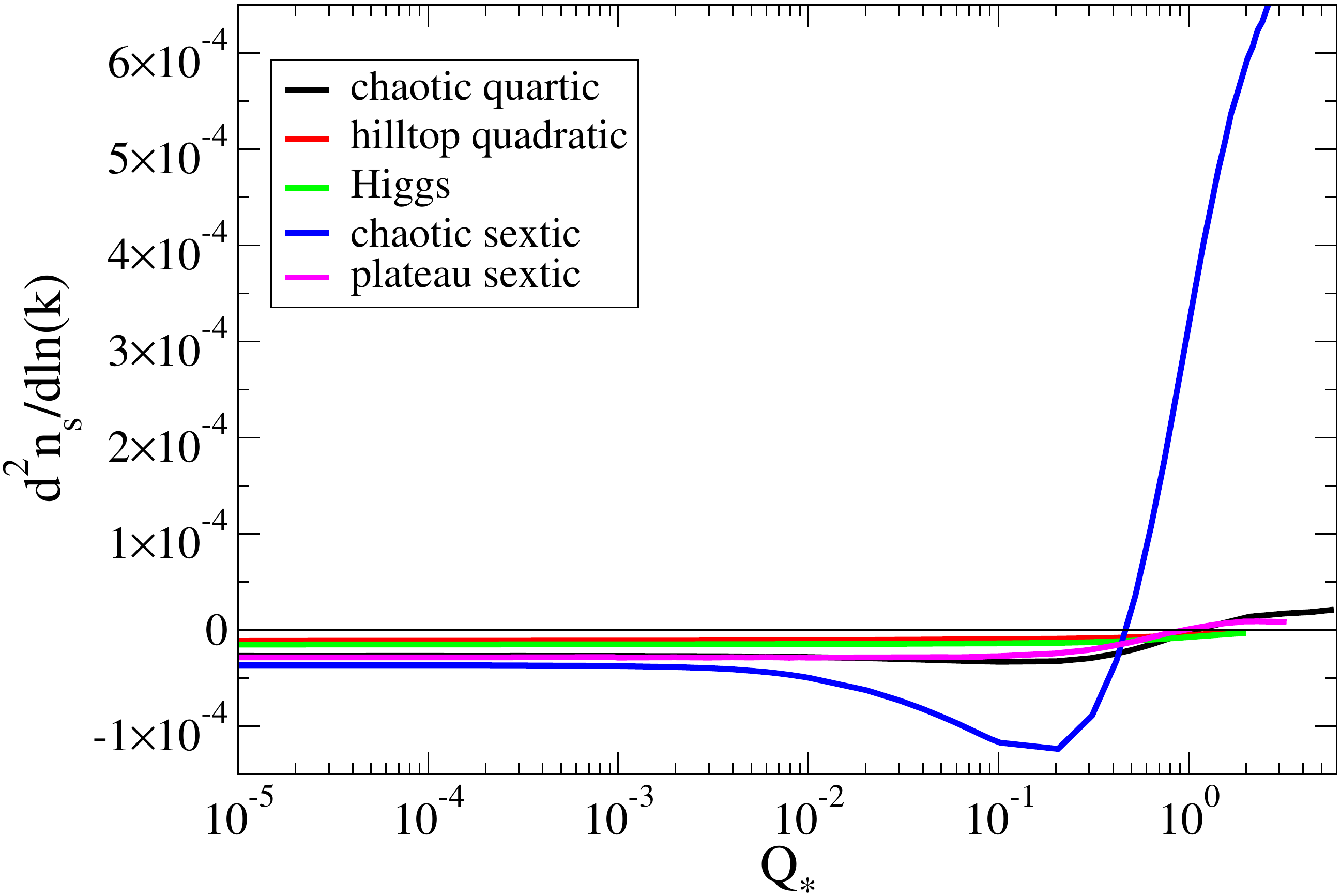}}
\caption{The running of the running, $n_s''$, as a function of the
  dissipation ratio $Q_*$.  Panel (a) are the results for the cubic
  dissipation coefficient, while panel (b)  gives the results when the
  dissipation coefficient is linear. }
\label{figd2ns}
\end{figure}
\end{center}
%%%%%%%%%%%%%%%%%%%%%%%%%%%%%%%%%%%%%%%%%
smaller values of the dissipation ratio and become positive for $Q_* \gtrsim
0.01$ when we assume a cubic dissipation regime,
or $Q_*\gtrsim 1$ for the linear dissipation case.  
Also for the Higgs potential the $n_s'$ turns positive for values of $Q_*$ 
that lie beyond the contours of the $(n_s,r)$ plane of 
{}Fig.~\ref{fignsXr}, which are out of our interest.
The behaviors shown in Fig.~\ref{figdns}
are very remarkable since the current Planck data constrain a tiny
negative value for the running at $68\%$ CL when the extension with
$n_s'$ to the minimal model is considered,  while prefer a tiny
positive value when also $n_s''$ is assumed.  In the CI model only
negative values for the running are predicted, on the contrary in WI
we are observing more possibilities. 

In {}Fig.~\ref{fig:2D_run}, the left panels always show the
predictions for the cubic dissipation regime, while in the right
panels  are the results for the linear dissipation case. 
In panels (a) and (b) we plot the WI theoretical predictions compared with the two-dimensional confidence region of the  ($n_s'$, $n_s$) plan for the $\Lambda$CDM$+n_s'$ model (yellow contours).
The remaining (c)-(d) panels %of {}Fig.~\ref{fig:2D} 
show the predictions for the $\Lambda$CDM$+n_s'+n_s''$ model (cyan contours) from Planck TT+lowP data.  
We note that the running $n_s'$ WI predictions are 
%
%%%%%%%%%%%%%%%%%%%%%%%FIG 5%%%%%%%%%%
\begin{center}
\begin{figure}[!htb]
\subfigure[]{\includegraphics[width=4cm]{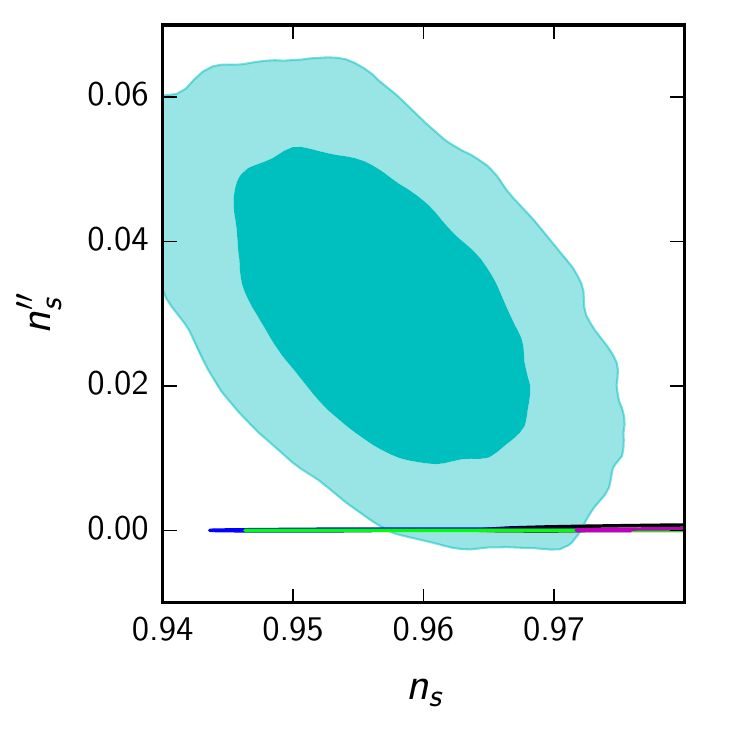}}
\subfigure[]{\includegraphics[width=4cm]{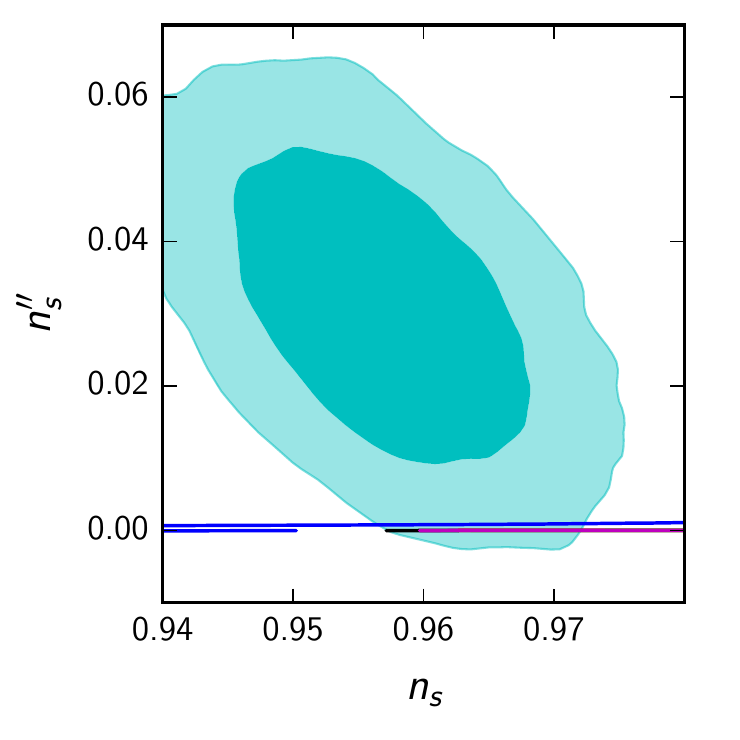}}
\caption{Marginalized joint $68\%$ and $95\%$ C.L. for the
  $\Lambda$CDM$+n_s'+n_s''$ model from Planck TT+lowP
  data.  The CL are compared with the WI theoretical predictions of
  the cubic dissipation regime (left column) and linear dissipation
  regime (right column).  The color schemes for the curves (not easily
  distinguishable in the scale of the plots) refer to each of the
  inflaton potentials considered and are the same as the ones used in
  {}Fig.~\ref{figdns}.}
\label{fig:2D_runrun}
\end{figure}
\end{center}
%%%%%%%%%%%%%%%%%%%%%%%%%%%%%%%%%%%%%%%%%
inside the $68\%$ CL, depending from the $Q_*$ value, for both the analysis.  

We also show the running of the
running  $n_s''$ as a function of $Q_*$ in {}Fig.~\ref{figd2ns}, when are considered the cubic and linear dissipation coefficients, respectively
panels (a) and (b). Similarly to what happened for the
running, also for the running of the running  all the models starts
with a small negative value, which then for some larger value of $Q_*$
become positive. Also in this case the Higgs potential becomes
positive for much larger values of $Q_*$, not shown in the plots.
The behaviours seen for $n_s''$ are very peculiar, since the results given by the CI single field models, as well as 
non-interacting multifield inflaton models, are naturally not capable of explaining a positive running of the running.  In particular, all the models considered here predict in a CI context a negative $n_s''$ (see, e.g., Ref.~\cite{Escudero:2015wba}).   In the
WI picture, however, our results show that we can achieve  a positive
$n_s''$, in accordance to what is indicated by the more recent CMB
analysis~\cite{Planck2015,Cabass:2016ldu} (see also
Ref.~\cite{vandeBruck:2016rfv} for a discussion of an  alternative way
of producing a positive $n_s''$ in a somewhat intricate two-field CI
model).  
Noteworthy, the predicted value of $n_s''$ of {}Fig.~\ref{figd2ns} 
are very tiny, almost close to zero. 
In {}Fig.~\ref{fig:2D_runrun} we show the marginalized CL for the $\Lambda$CDM$+n_s'+n_s''$ model compared with the WI theoretical predictions of the cubic dissipation regime,
panel (a) and linear dissipation regime, panel (b). 
We can see that several analyzed values of $Q_*$ lie inside the $95\%$ CL.

\begin{center}
\begin{figure}[!htb]
\subfigure[]{\includegraphics[width=7.5cm]{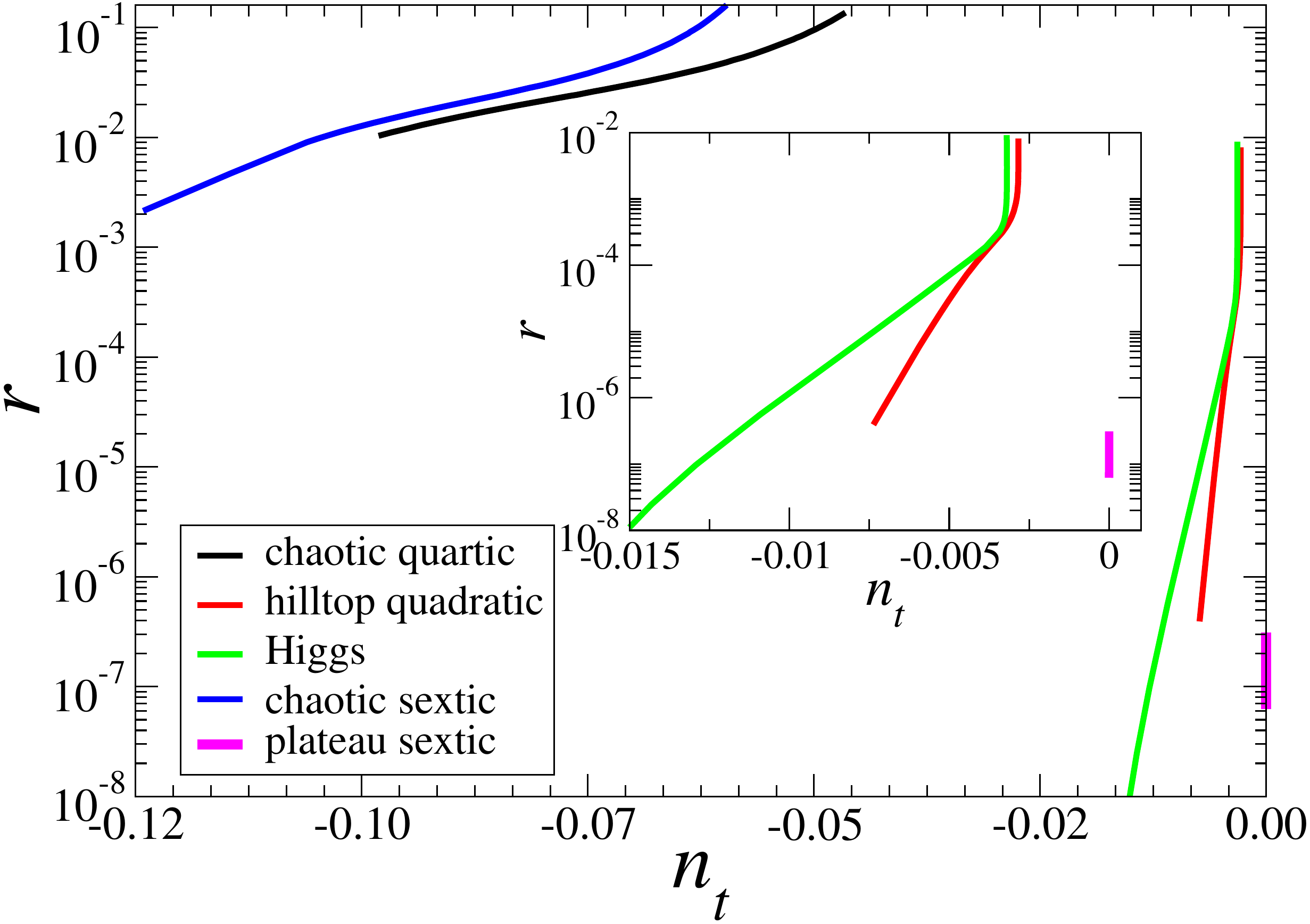}}
\subfigure[]{\includegraphics[width=7.5cm]{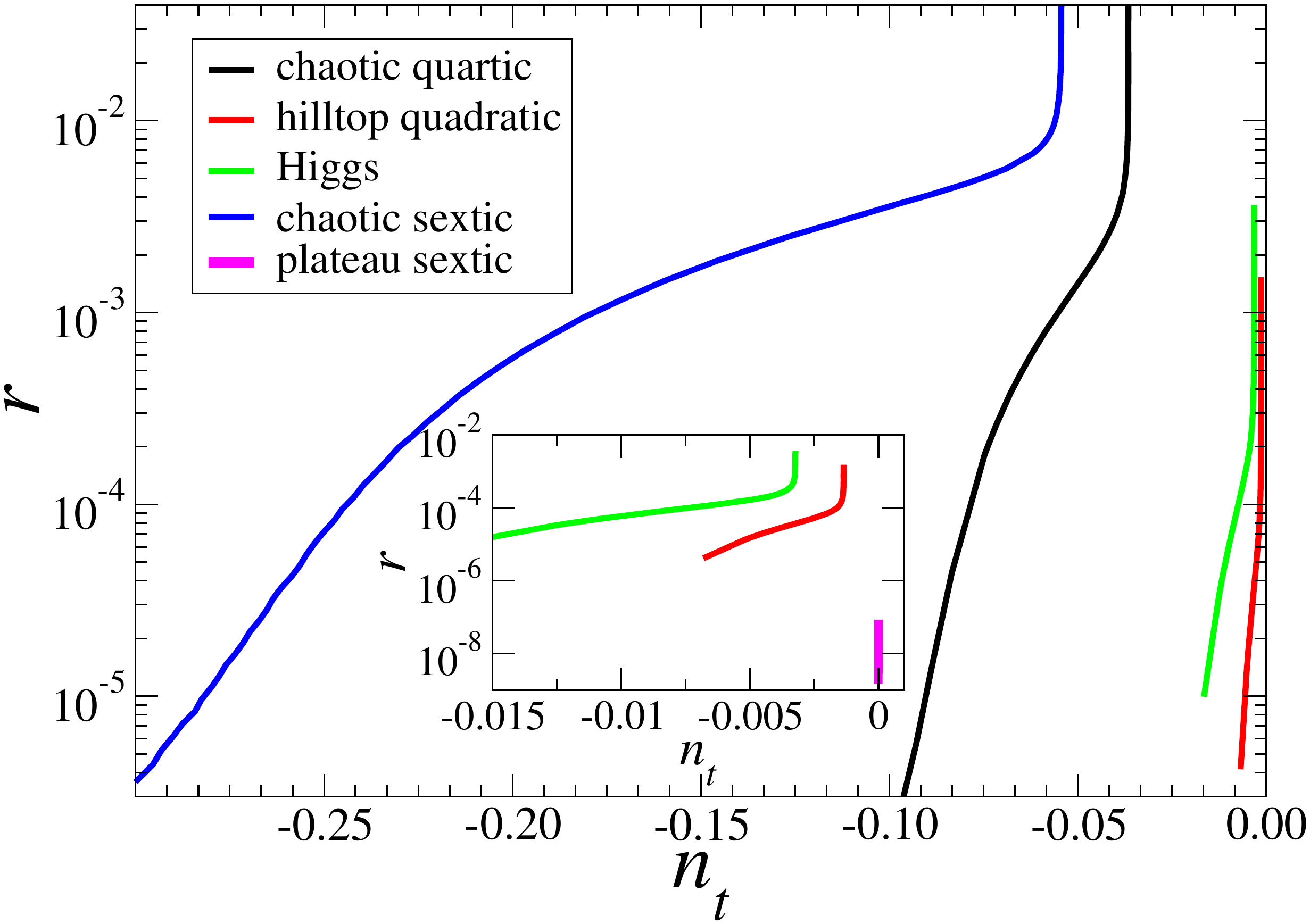}}
\caption{The tensor tilt $n_t$ and the tensor-to-scalar ratio $r$ in
  the plane $(n_t,r)$ for different values of the dissipation ratio
  $Q_*$.  Panel (a) are the results for the cubic dissipation
  coefficient, while panel (b) gives the results when the dissipation
  coefficient is linear.  {}For both cases of dissipation, the plateau
  sextic potential (magenta line) have a too small $r \lesssim
  10^{-8}$ and very small $n_t$ and whose  results are barely seen in
  the scale of the figures.}
\label{figntXr}
\end{figure}
\end{center}

{}For completeness, in {}Fig.~\ref{figntXr} we show the running $n_t$
of the tensor spectrum, $\Delta_T(k) \propto k^{n_t}$, as a function
of the tensor-to-scalar ratio $r$.  The dissipation ratio $Q_*$ increases from
$10^{-5}$ up the values shown in {}Fig.~\ref{fignsXr} and, 	consequently, 
the values of $r$ decrease for each potential treated.  We already noticed
that, in the WI context, the dissipation affects strongly the scalar
curvature power  
%%%%%%%%%%%%%%%%%%FIGURE06%%%%%%%%%%%%%%%%%%%
%%%%%%%%%%%%%%%%%%%%%%%%%%%%%%%%%%%%%%%%%
spectrum while the tensors remain unaffected.  Indeed,
the WI only indirectly affects the tensors through the change of
dynamics for the Hubble parameter, due to the presence of the radiation
bath.  Note that the consistency relation observed in CI,  $r=-8 n_t$,
is modified in the WI picture. It become $r< 8 |n_t|$, again as a
consequence of the enhancement factor Eq.~(\ref{calF}) that affects WI
but not CI. 

The value of $r$ decreases for both the regimes
with respect to the cold inflation, and this allows to a
larger range of parameter values (temperature and dissipation) to
enter in accordance with the Planck data.  In this sense, the WI induces
an increased degeneracy in the results for $r$ and $n_s$ for different
types of inflaton potentials. The degeneracy can in principle be
broken when we combine these results with those obtained for the other
observables like the running and the running of the running of the
spectral index, but these constraints are still 
%
%
%%%%%%%%%%%%%%%%%%%% TAB %%%%%%%%%%%%%%%%%%%%
\begin{table*}[!htb]
\caption{The values of dissipation ratio $Q_*$ and $T_*/H_*$,
  along also the values of $r$, $n_s'$ and $n_s''$, and the 
$\Delta \chi^2$ with respect to the minimal $\Lambda$CDM model
  when $n_s$ is  fixed at the value $n_s \simeq 0.9655$, 
  for each model considered in this work.}
\begin{center}
\begin{tabular}{c|c|c|c|c|c|c|c}
  \hline  \hline $V(\phi)$        & $\Upsilon$                  &
  $Q_*$   & $T_*/H_*$  & $r$  & $n_s^\prime$ & $n_s^{\prime\prime}$ & $\Delta \chi^2_{\rm min}$
  \\ 
\hline 
quartic  &    &  $1.697  \times 10^{-3}$ & $7.246$  & $0.036$   & $-9.840 \times  10^{-4}$
  & $-2.557 \times 10^{-5}$   & -0.2
\\ 
sextic  &  &  $0.187$ & $41.945$  & $5.225 \times 10^{-3}$    & $1.540 \times
  10^{-3}$   & $1.972 \times 10^{-4}$   & +0.3
\\   
hilltop &  $\propto\frac{T^3}{\phi^2}$ &  $0.186$ & $41.656 $  & $1.741 \times
  10^{-4}$ & $-9.997 \times 10^{-5}$  & $-3.101 \times 10^{-6}$  & +0.1
\\    
Higgs  &   & $1.417$ & $214.829$ & $2.317 \times 10^{-6}$  & $-1.857 \times
  10^{-4}$ & $-4.333 \times 10^{-6}$  & -0.1
\\  
plateau sextic   &  &  $5.645 \times 10^{-3}$ & $10.766$ & $1.085 \times 10^{-7}$  &
  $-4.692 \times 10^{-4}$ & $3.369 \times 10^{-5}$  & 0
\\  
\hline  quartic  &   &  $1.256$ & $273.472$ & 0
  $1.27639 \times 10^{-4}$ & $-3.019 \times 10^{-4}$  & $4.156 \times
  10^{-6} $  & 0
\\    
sextic  &  &  $4.966$ & $769.074$ & $1.064 \times 10^{-5}$ & $3.731 \times
  10^{-3}$  & $8.381 \times 10^{-4}$ & +0.7
\\ 
hilltop  & $\propto T$  &  $0.028 $ & $50.303 $  & $1.092 \times 10^{-4}$   & $-1.58837
  \times 10^{-4}$  & $-1.028 \times 10^{-5} $ & 0
\\  
Higgs   &  &  $0.020$ & $44.492$  & $2.947 \times 10^{-4}$  & $-2.50462 \times
  10^{-4}$  & $-1.424 \times 10^{-5}$ & -0.1
\\    
plateau sextic  &  &  $0.810$ & $210.187$  & $4.448 \times 10^{-9}$  & $-3.862 \times
  10^{-4}$  & $-2.708 \times 10^{-6}$ & 0
\\    
\hline
\end{tabular}
\end{center}
\label{tab:models} 
\end{table*}
%
%
%%%%%%%%%%%%%%%%%%%%%%%%%%%%%%%%%%%%%%%%%%%%%%%%
too small and can remain within the Planck allowed ranges. 
\section{Analysis and Results} 
\label{sec:analysis and results}
Let us now study the statistical viability of the WI cases of
inflaton models considered in the previous section, for both the
dissipation regimes defined by  Eqs.~(\ref{upsilon}) and
(\ref{upsilon2}), using the most up-to-date CMB data.
To perform our analysis, we consider a minimal $\Lambda$CDM model and
modify the standard primordial power-law spectrum following the
equations for the WI picture.  Therefore, we vary the usual
cosmological parameters, namely, the physical baryon density,
$\Omega_bh^2$, the physical cold dark matter density, $\Omega_ch^2$,
the ratio between the sound horizon and the angular diameter distance
at decoupling, $\theta$, and the optical depth, $\tau$. We do not use
the primordial parameters $\mathcal A_s$ and $n_s$, respectively the
scalar amplitude and the spectral index, since we parameterise the
primordial spectrum as in Eq.~(\ref{Pk}).  Both $P_0(k/k_*)$ and
${\cal F}(k/k_*)$ are obtained numerically for the studied potentials
by solving the background equations in WI for different values of the
the dissipation ratio $Q_*$.  These values are calculated for the
scales leaving the Hubble radius in an interval $\Delta N=5$ around
the value $N_*$, where the pivot scale crosses the horizon.  This
corresponds to an analysis of $10$ e-folds of inflation, which is
roughly the range of values of primordial expansion probed by the
current CMB data.  We note that the dissipation ratio $Q_*$, the 
temperature ratio $T_*/H_*$ and the amplitude $P_0(k/k_*)$ of 
Eq.~(\ref{Pk}) are found to be linear
with the scale for the considered potentials and for both the
dissipation regimes.  Hence, we can approximate them in our analysis
with a power-law fitting without loss of information.  Our
parameterization is proved to be very accurate, taking into account
higher order effects such as the running  and the running of the
running informations. In our analysis we also vary the nuisance foreground
parameters~\cite{Aghanim:2015xee} and consider purely adiabatic
initial conditions. The sum of neutrino masses is fixed to $0.06$ eV,
and we limit the analysis to scalar perturbations with $k_*=0.05$
$\rm{Mpc}^{-1}$. 

To obtain our results, we use a modified version of the {\sc
  CAMB}~\cite{camb}  to compute the theoretical CMB anisotropies
spectrum in a WI context for different values of the dissipation ratio
$Q_*$.  In order to compare these theoretical predictions with
observational data, we employ a Monte Carlo Markov Chain analysis via
the publicly available package {\sc CosmoMC}~\cite{Lewis:2002ah}.
{}Finally, to maximize the likelihood and write our results, we use
the Bound Optimization BY Quadratic Approximation (BOBYQA)
algorithm~\cite{Powell}, which is an optimized method implemented in
the {\sc CosmoMC} code in order to minimize the $\chi^2$ value of the
models.

We choose to use the second release of Planck data ``TT+lowP", namely
the high multipole Planck temperature data (TT)  from the 100-,143-,
and 217-GHz half-mission TT cross-spectra and the low-$\ell$ data by
the joint TT, EE, BB and TE likelihood (lowP)~\cite{Aghanim:2015xee}.
We choose to not include high-$\ell$ polarization data since they have
not a significant impact on the constraint of the $n_s'$ and $n_s''$
parameters, as already found by the Planck
Collaboration~\cite{Planck2015}.   Indeed, both these parameters
affect more the low multipoles than the other scales.  
We also decide to not include tensor data in our analysis. 
Since the $r$ values of all the models we are going to analyze
%produce observables with $r$ value 
are inside the current $68\%$ C.L. of the TT+lowP data, 
it is reasonable to expect that tensor data do not add significant information. 

As we saw in {}Fig.~\ref{fignsXr}, the dissipative ratio $Q_*$
variation produces a significant change in the value of $n_s$ for each
potential considered.  Since the data show very accurate constraint on
the $n_s$ parameter, we expect fine constraints also on $Q_*$.   At
the same time, the Planck data prefers $\simeq 0.9655$ as central value
for $n_s$, so it is easy to suppose that the most preferred dissipative
ratio is the one that guarantees that value.  In our analysis we
explore the wide range of $Q_*$ shown in {}Fig.~\ref{fignsXr}, but for
%%%%%%%%%%%%%%%%%
\begin{center}
\begin{figure}[!htb]
\subfigure[]{\includegraphics[width=8cm]{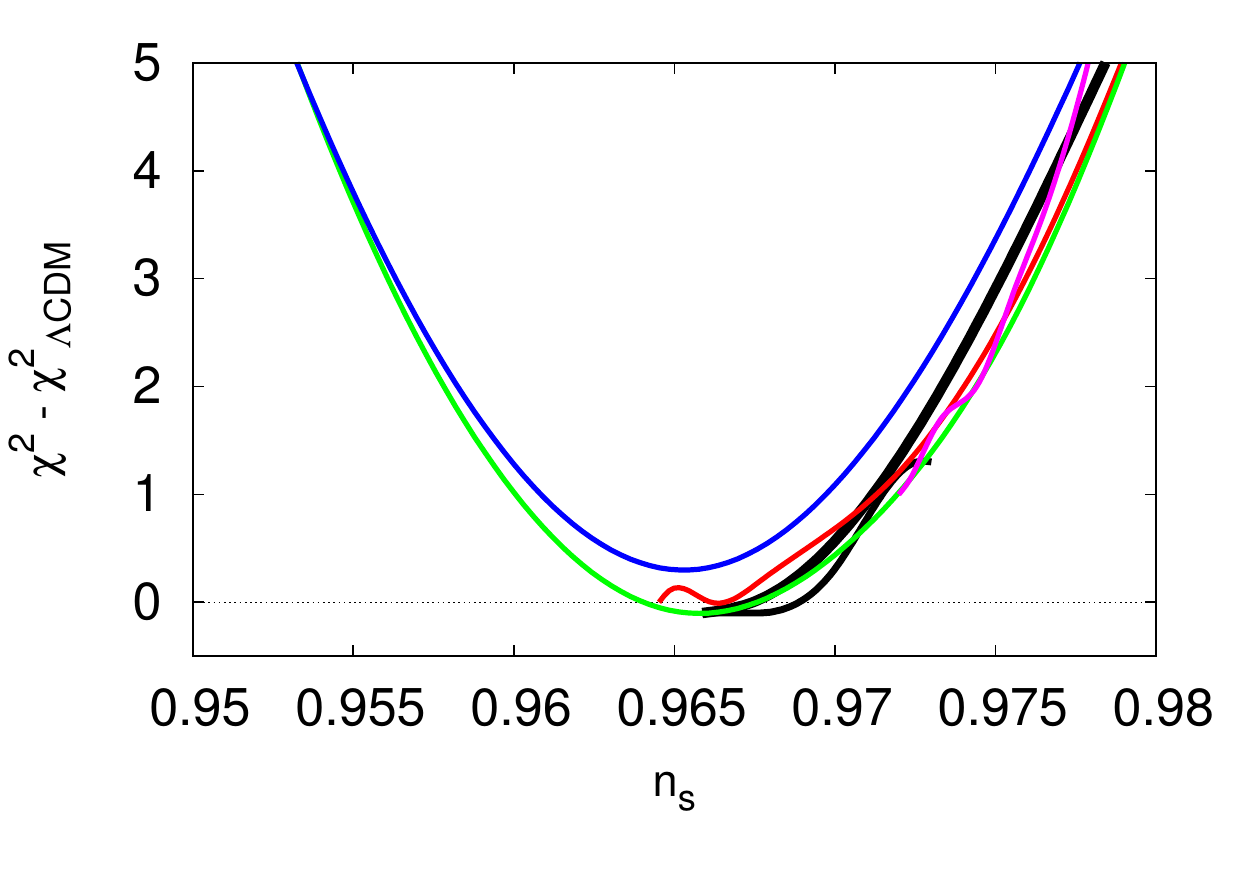}}
\subfigure[]{\includegraphics[width=8cm]{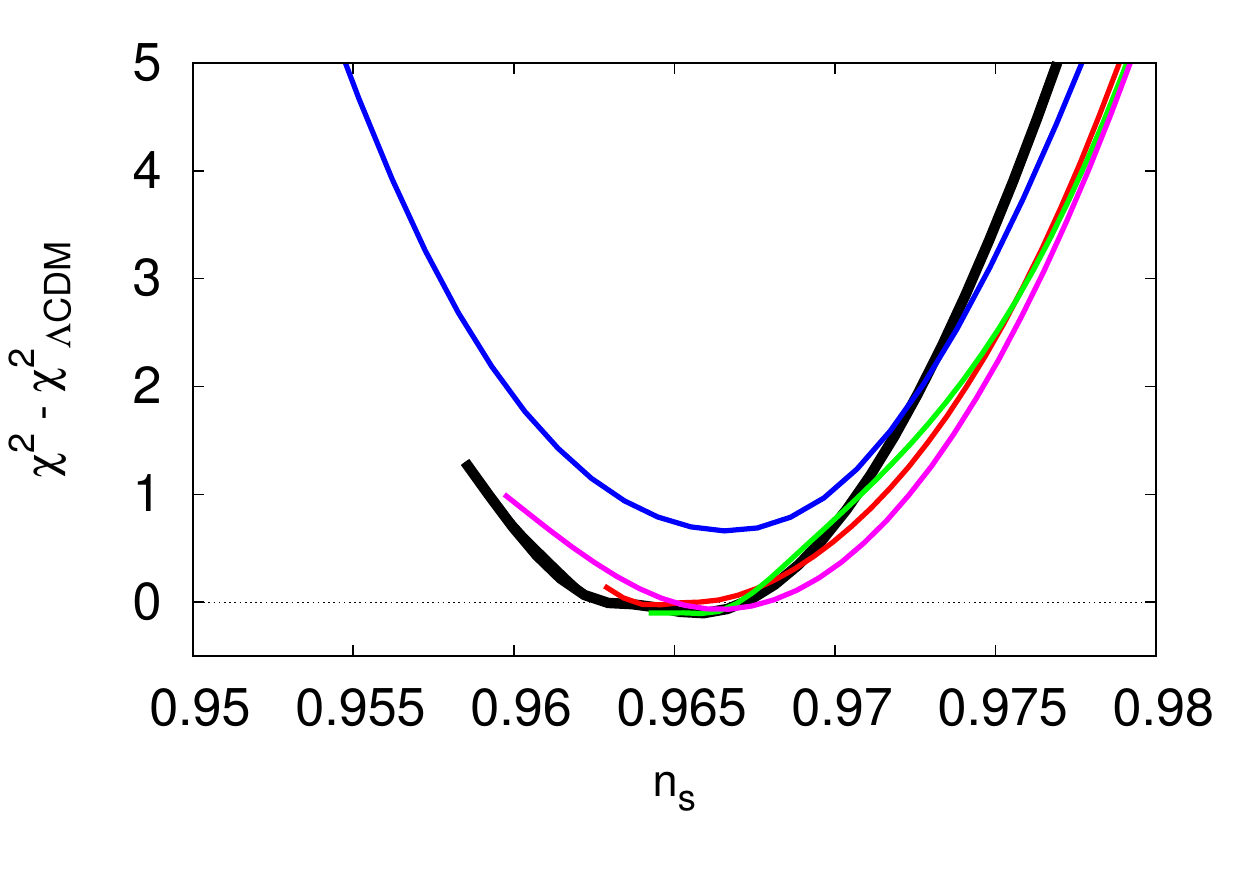}}
\caption{The behavior of the $\chi^2$ value as a function of the
  spectral index $n_s$ for the chaotic quartic potential (black line),
  hilltop (red line), Higgs potential (green line), chaotic sextic
  potential (blue line) and for the plateau sextic potential (magenta
  line). Panel (a): Results for the cubic dissipation
  coefficient. Panel (b): Results for the linear dissipation
  coefficient.}
\label{chi2_ns}
\end{figure}
\end{center}
%%%%%%%%%%%%%%%%%%%%%%%%%%%%%%%%%%%%%%%%
illustration we show in Tab.~\ref{tab:models} the observational
predictions  when $n_s$ is fixed at the expected central value $n_s
\simeq 0.9655$.  The value of the  $\Delta \chi^2$, with respect to
the minimal $\Lambda$CDM data, is also quoted for illustration.  We
stress that this value cannot correspond to the best-fit value of the
model, though it  is still met for very close value. We note that the
chaotic sextic model is the only one that produces,  for the favorite
$n_s$ value, a positive $n_s'$ and $n_s''$.  At the same time, it
shows the worst $\Delta \chi^2$.  This is due to the very small
contribution of the running of the  running, lesser than $10^{-4}$
for all the WI models.
%,  and we stress that it is almost unseen by the data in our analysis.  
Indeed, as we just saw in the previous section 
for the $\Lambda$CDM+$n_s'$ analysis, the data show a preference for negative value of the running when $n_s''$ is zero (or close to zero). 

\begin{center}
\begin{figure}[!htb]
\subfigure[]{\includegraphics[width=8cm]{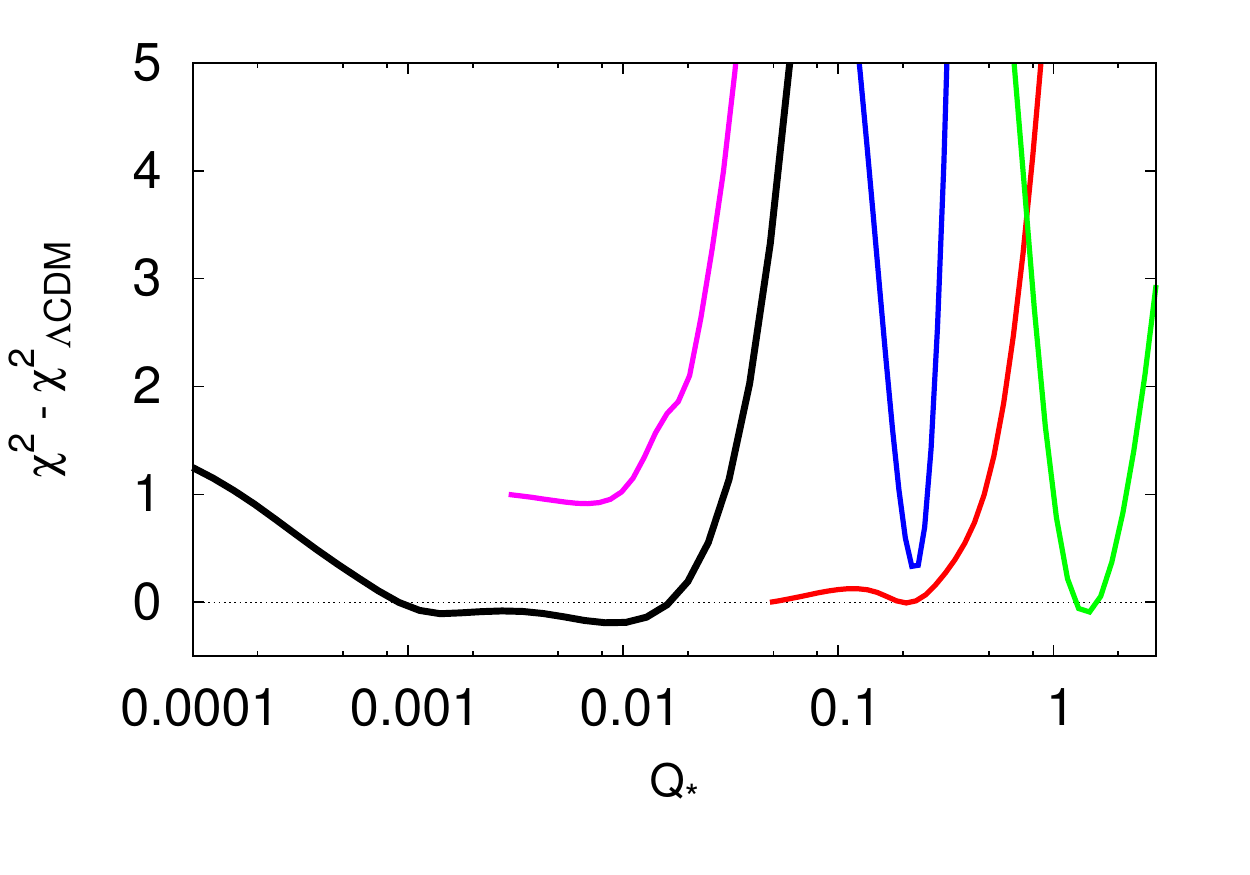}}
\subfigure[]{\includegraphics[width=8cm]{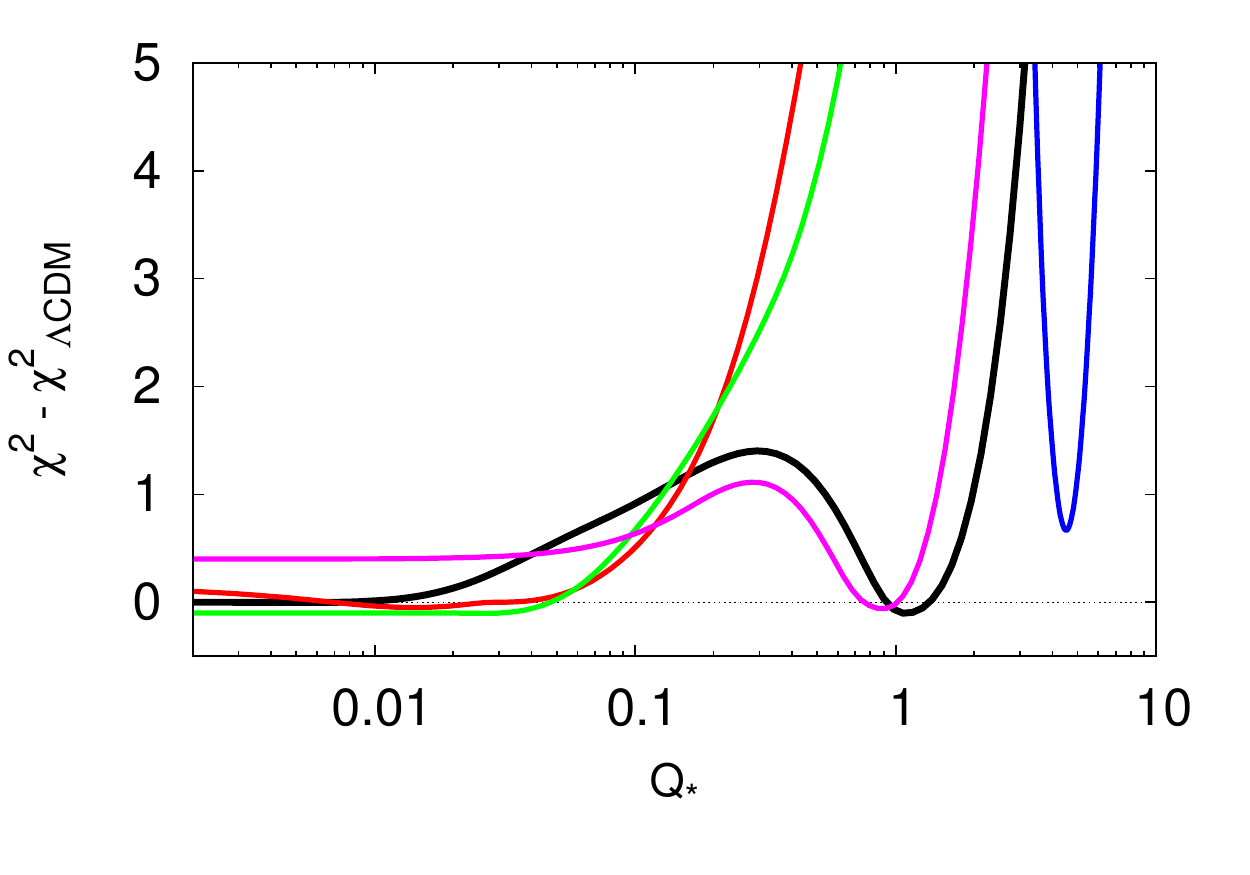}}
\caption{The behavior of the $\chi^2$ value as a function of the
  dissipation ratio $Q_*$ for the chaotic quartic potential (black
  line), hilltop (red line), Higgs potential (green line), chaotic
  sextic potential (blue line) and for the plateau sextic potential
  (magenta line). Panel (a): Results for the cubic dissipation
  coefficient. Panel (b): Results for the linear dissipation
  coefficient.}
\label{chi2_Q}
\end{figure}
\end{center}

The results of our analysis are presented in {}Figs.~\ref{chi2_ns} and
~\ref{chi2_Q}, where we show the $\Delta \chi^2$ values of each
considered 
%
%%%%%%%%%Fig8%%%%%%%%%%
%%%%%%%%%%%%%%%%%%%%%%%%%%%%%%%%%%%%%%%%%%%%%%%%
%
model, always with respect to the minimal $\Lambda$CDM model, as a function of the spectral index $n_s$
and in terms of the dissipative ratio $Q_*$, respectively.
As before, in the panels (a) we show
the results assuming the cubic dissipation, while in the panels (b)
are the results for the linear dissipation case.   In the
{}Fig.~\ref{chi2_ns} the variation of $\chi^2$ is of order of unity
when the $n_s$ parameter value varies within its $1 \sigma$ error,
while rapidly increases until $\Delta \chi^2 \sim 5$, for $n_s$
varying within $95\%$ CL.  We also note that the minimum best fit of
these WI models is compatible with the $\Lambda$CDM one.
Noteworthy, the chaotic sextic model shows the worst $\Delta\chi^2$,
which is seen more pronounced in the linear dissipation regime.   
The reason becomes clear looking the curves of this model in the panel (b) of
{}Figs.~\ref{fignsXr} and \ref{figdns}, where we can note that only  values
of $3 < Q* < 6.5$ reconcile the $n_s$ into the $95\%$ C.L. of the
Planck data, 
%
%%%%%%%%%%%%%%%%%FIGURE 9%%%%%%%%%%%%%%%%%%%
\begin{center}
\begin{figure*}[!htb]
\subfigure[]{\includegraphics[width=15cm]{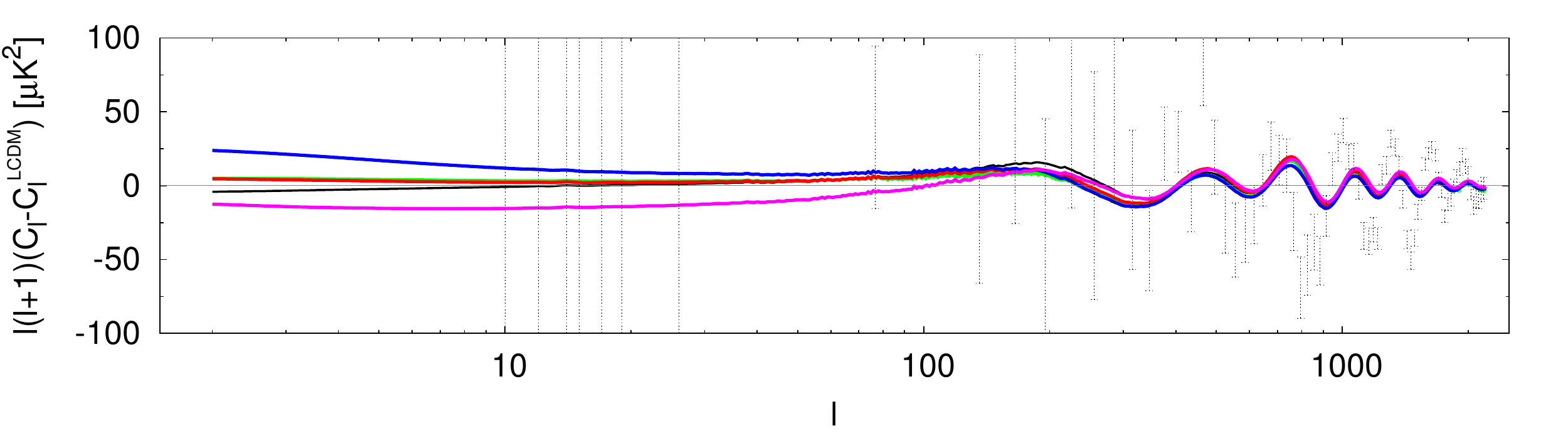}}
\subfigure[]{\includegraphics[width=15cm]{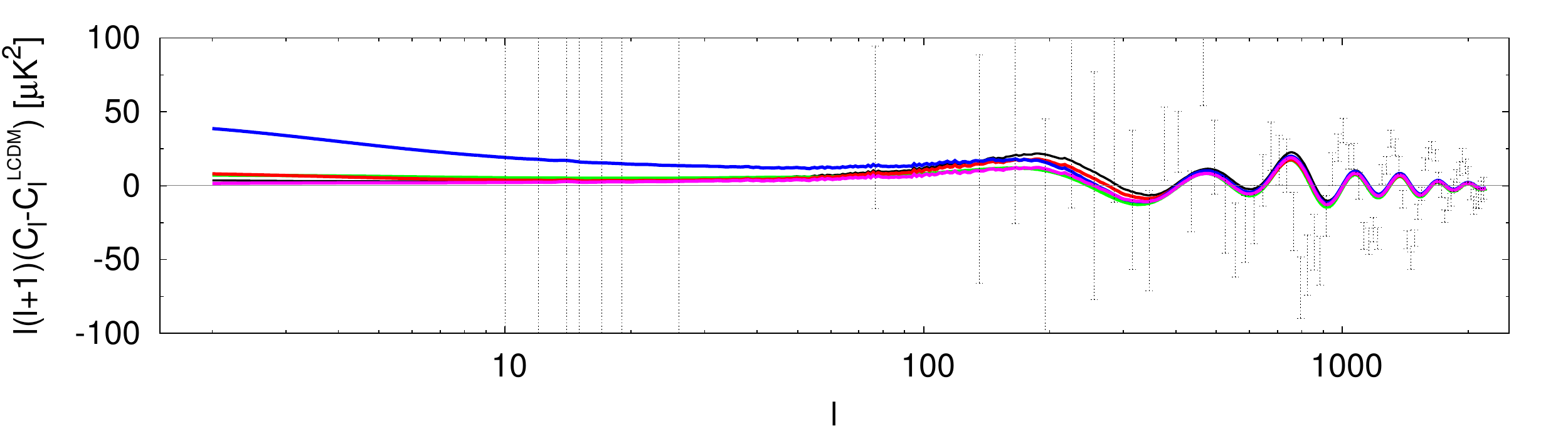}}
\caption{Temperature power spectrum residual plots for
  the best fit of the chaotic quartic potential (black line), hilltop
  (red line), Higgs potential (green line), chaotic sextic potential
  (blue line) and for the plateau sextic potential (magenta line) with
  respect to the $\Lambda$CDM model (grey line).  
  Data shown are the TT+lowP (Planck 2015). 
  Panel (a): Results
  for the cubic dissipation coefficient. Panel (b): Results for the
  linear dissipation coefficient.}
\label{fig:best-fit_TT}
\end{figure*}
\end{center}
%
%%%%%%%%%%%%%%%%%%%%%%%%%%%%%%%%%%%%%%%
but for values of $Q*>1$ it shows positive running,
which is ruled out by the same data for the $\Lambda$CDM+$n_s'$
analysis.  The latter is assumed as reference since, as we have
already mentioned above, for these WI models the  running of the
running $n_s''$ is too small and the data accuracy is not enough to
detect the tiny signal.  The incompatibility to satisfy both the
observables causes the worse $\Delta\chi^2$.   At the same time we
note that there are values of $Q_*$, for all the other potentials, that
allow combinations of the $n_s$ and its higher order parameters such that all the current observations are fulfilled.
{}Finally, in {}Fig.~\ref{chi2_Q} we can see that the  significantly
different behavior between the cubic and linear dissipation cases is
reflected in a further preference for higher values of $Q_*$ in the latter for all the models, except for the hilltop and Higgs models.

The differences between the best fit temperature power spectrum 
of our analysis and the minimal $\Lambda$CDM model is shown in
{}Fig.~\ref{fig:best-fit_TT}. Here, the best fit WI models are very
close to the minimal $\Lambda$CDM predictions, with slightly different
amplitude with respect to the simplest power-law potential.  Our work
shows that these slight variations are too small to be appreciated 
by the current observations with a $\chi^2$ statistical analysis. 
At the same time, the data well constrain
the observables determined by the dissipation ratio value and we
obtained the first estimate of how $Q_*$ affects the anisotropy
temperature spectra.  Also, we can comment that the two dissipation
coefficient forms considered do not produce specific features in the
spectrum.  As already mentioned, we want to stress that as our WI
parameterization takes into account one degree of freedom less than the
$\Lambda$CDM model and ensures a best fit $\Delta\chi^2 \sim 0$.

{}Finally, it is interesting to quote the current best fit value, in the
CI context, of several inflaton potentials we have analyzed
~\cite{Planck2015}.  {}For instance, for the chaotic quartic potential
model it was estimated $\Delta\chi^2= +43.3$ with respect to the
$\Lambda$CDM model; for the hilltop potential $\Delta\chi^2 = +4.4$
and for the Higgs potential $\Delta\chi^2 = +5.5$.  These values can
be compared with the $\Delta\chi^2$ values shown in the
Tab.~\ref{tab:models}, or in {}Figs~\ref{chi2_ns} and \ref{chi2_Q}.
Then, our analysis gives a good indication of how the WI picture
reconciles the different inflaton potentials considered, producing
observables in agreement with the data.
%
%%%%%%%%%%%%%%%%%%%%%%%%%%%%%%%%%%%%%%%%%%%%%%%%%%%%%%%%%%%%%
\section{Conclusions}
\label{sec:Conclusions}
In this work we performed extended analysis of several representative
inflaton  potentials in a warm inflation scenario, covering both large
and  small field models. 
We obtain our results studying two different dissipation regimes, namely with the dissipation coefficient showing a cubic and linear dependence with the temperature of the thermal bath.

The dynamics leading to the dissipation coefficient with a cubic form 
emerges in what is usually
referred as the {\it low temperature} regime of WI,  where the
inflaton is coupled to heavy intermediate fields that are in turn
coupled to light radiation fields.  The decay of the heavy
intermediate fields into the light radiation fields produces 
the cubic dependence of the dissipation coefficient.
~\cite{Berera:2008ar,Bastero-Gil2011,BasteroGil:2012cm}.
The second dissipation regime we have considered is known in the literature as the {\it high temperature} regime of WI.  
Here the inflaton field directly couples to the radiation fields, 
yet  preventing large thermal and radiative corrections to the inflaton potential, 
as a consequence of a collective symmetry breaking in the model construction.  
This leads to a dissipation coefficient that is linear in the temperature
~\cite{Bastero-Gil:2016qru}.

We achieve a parameterization of the primordial power spectrum able to
take into account higher order effects as the running and the running of the running, and we analyze the
observational predictions of the considered models in the two dissipation regimes described above.  
We show that the tensor modes are not affected by
the dissipative dynamics, while the  scalar spectra is modified by an
additional enhancement term.  Hence, the dissipation ratio $Q_*$ 
shows a significant impact on
the spectral index $n_s$ and the tensor-to-scalar ratio $r$.  
It can reconcile the parameters value
with the observation, i.e., a non detection  of primordial gravity
waves might be a hint of WI in the high dissipative regime ($Q_*
\gtrsim 1$).  In other words, a change of $Q_*$
allows a larger range of cosmological parameter values 
to enter in accordance with the Planck results.  
This degeneracy can in principle be broken combining these results with the constraints on the other observables.
Other way is to study the non-Gaussianities, since in WI
these have distinct features when compared with the CI
picture~\cite{Bastero-Gil:2014raa}.  We can also distinguish the
regimes of weak dissipation  $Q_* \ll 1$, with those of strong
dissipation $Q_* \gtrsim 1$ through different  shapes, unique to WI
(for details, see, e.g., Ref.~\cite{Bastero-Gil:2014raa}). 

Intriguing, our studies show that both the running $n_s'$ and the running of the running $n_s''$  assume
negative or positive values depending on $Q_*$.  These behaviors are
in contrast with the CI results, where only  negative values for
$n_s'$ and $n_s''$ are predicted.  The WI picture is demonstrating a
better versatility, where the parameter $Q_*$ (or, equivalently, the
temperature of the thermal bath present during the WI dynamics) is the
key parameter,  and can achieve values able to better explain the current data.  

We compared the predictions with the most up-to-date CMB data and we
performed a statistical analysis to constrain the dissipative
effects.  The data  well constrain the observables determined by the
dissipation ratio value, and we get the first estimate of how
$Q_*$ affects the anisotropy temperature power spectrum.  Our results show
the agreement between the WI picture with the Planck release,
rehabilitating several forms of primordial potentials ruled out by the
data in the CI context.  At the same time, our analysis show that the
chaotic sextic model is unable to describe the current data,  since there
is not a value of $Q_*$ that is able to satisfy simultaneously the values
required for the spectral index and its running.  
Despite the values for $n_s''$ obtained for all the models are still to small to be of relevance, we note that its magnitude tends
to increase with the gain in dissipation value. 
It is a feature that can eventually be
explored in future model building in WI.  
At the same time, the
current constraints on the higher order parameters are still too wide, improvements in the data are need to update our results.

%%%%%%%%%%%%%%%%%%%%%%%%%%%%%%%%%%%%%%%%%%%%%%%%%%%%
\acknowledgments  M.~B. acknowledges financial support from the
Funda\c{c}\~ao Carlos Chagas Filho de Amparo \`a Pesquisa do Estado do
Rio de Janeiro - FAPERJ (\textit{fellowship Nota 10}).   R.~O.~R.~is
partially supported by  Conselho Nacional de Desenvolvimento
Cient\'{\i}fico e Tecnol\'ogico - CNPq (Grant No. 303377/2013-5) and
%Funda\c{c}\~ao Carlos Chagas Filho de  Amparo \`a Pesquisa do Estado
%do Rio de Janeiro - 
FAPERJ (Grant No. E - 26 / 201.424/2014).  The
authors acknowledge the use of the  CosmoMC
code~\cite{camb,Lewis:2002ah}.
%%%%%%%%%%%%%%%%%%%%%%%%%%%%%%%%%%%%%%%%%%%%%%%%%%%%

%%%%%%%%%%%%%%%%%%%%%%%%%%%%%%%%%%%%%%%%%%%%%%%%%%%%%%          

\end{document}